\documentclass[aps,preprint]{revtex4}%
\usepackage{amsfonts}
\usepackage{amsmath}
\usepackage{amssymb}
\usepackage{graphicx}%
\providecommand{\U}[1]{\protect\rule{.1in}{.1in}}

\begin{document}
\title{Precision benchmark calculations for four particles at unitarity}
\author{{Shahin Bour$^{a}$, Xin Li$^{b}$, Dean~Lee$^{b}$, Ulf-G.~Mei{\ss }ner$^{a,c}$,
Lubos Mitas$^{b}$}}
\affiliation{$^{a}$Helmholtz-Institut f\"{u}r Strahlen- und Kernphysik (Theorie) and
\linebreak Bethe Center for Theoretical Physics, Universit\"{a}t Bonn, D-53115
Bonn, Germany \linebreak$^{b}$Department of Physics, North Carolina State
University, Raleigh, NC 27695, USA \linebreak$^{c}$Institut f\"{u}r Kernphysik
(IKP-3), Institute for Advanced Simulation (IAS-4), and J\"{u}lich Center for
Hadron Physics, Forschungszentrum J\"{u}lich, D-52425 J\"{u}lich, Germany}

\begin{abstract}
The unitarity limit describes interacting particles where the range of the
interaction is zero and the scattering length is infinite. \ We present
precision benchmark calculations for two-component fermions at unitarity using
three different \textit{ab initio} methods: \ Hamiltonian lattice formalism
using iterated eigenvector methods, Euclidean lattice formalism with
auxiliary-field projection Monte Carlo, and continuum diffusion Monte Carlo
with fixed and released nodes. \ We have calculated the ground state energy of
the unpolarized four-particle system in a periodic cube as a dimensionless
fraction of the ground state energy for the non-interacting system. \ We
obtain values $0.211(2)$ and $0.210(2)$ using two different Hamiltonian
lattice representations, $0.206(9)$ using Euclidean lattice, and an upper
bound of $0.212(2)$ from fixed-node diffusion\ Monte Carlo. \ Released-node
calculations starting from the fixed-node result yield a decrease of less than
$0.002$ over a propagation of $0.4E_{F}^{-1}$ in Euclidean time, where $E_{F}$
is the Fermi energy. \ We find good agreement among all three \textit{ab
initio} methods.

\end{abstract}

\maketitle

\section{Introduction}

The unitarity limit describes interacting particles where the range of the
interaction is zero and the S-wave scattering length is infinite. \ In this
paper we consider the unitarity limit of two-component fermions. \ Throughout
our discussion we refer to the two degenerate components as up and down spins,
though the correspondence with actual spin is not necessary. \ At sufficiently
low temperatures the spin-unpolarized system is an S-wave superfluid with
properties in between a Bardeen-Cooper-Schrieffer (BCS) fermionic superfluid
at weak coupling and a Bose-Einstein condensate of dimers at strong coupling
\cite{Eagles:1969PR,Leggett:1980pro,Nozieres:1985JLTP}. \ In nuclear physics
the phenomenology of the unitarity limit approximately describes cold dilute
neutron matter. \ The scattering length for elastic neutron-neutron collisions
is about $-18$~fm while the range of the interaction is roughly the Compton
wavelength of the pion, $1.4$~fm. \ The unitarity limit is approximately
realized when the interparticle spacing is about $5$ fm. \ While these
conditions cannot be produced experimentally, neutrons at around this density
can be found in the inner crust of neutron stars. \ 

Experimental probes of the unitarity limit are now well established using
trapped ultracold Fermi gases of alkali atoms. \ The characteristic length
scale for the interatomic potential is the van der Waals length $\ell
_{\text{vdW}}$. \ In the dilute limit the spacing between atoms can be made
much larger than $\ell_{\text{vdW}}$ and the interatomic potential is well
approximated by a zero-range interaction. \ The S-wave scattering length can
be tuned using a magnetic Feshbach resonance
\cite{O'Hara:2002,Gupta:2002,Regal:2003,Bourdel:2003,Gehm:2003}. \ This
technique involves setting the energy level for a molecular bound state in a
\textquotedblleft closed\textquotedblright\ hyperfine channel to cross the
scattering threshold for the \textquotedblleft open\textquotedblright%
\ channel. \ The total magnetic moments for the two channels are different,
and so the crossing can be produced using an applied magnetic field.

The ground state for two-component fermions in the unitarity limit has no
physical length scales other than the average distance between particles.
\ The scaling properties in the unitarity limit are the same as that of a
non-interacting Fermi gas. \ For $N_{\uparrow}$ up spins and $N_{\downarrow}$
down spins in a given volume we write the energy of the unitarity-limit ground
state as $E_{N_{\uparrow},N_{\downarrow}}^{0}$. \ For the same volume we call
the energy of the free non-interacting ground state $E_{N_{\uparrow
},N_{\downarrow}}^{0\text{,free}}$. \ In the following we write the
dimensionless ratio of the two energies as $\xi_{N_{\uparrow},N_{\downarrow}}%
$,%
\begin{equation}
\xi_{N_{\uparrow},N_{\downarrow}}=E_{N_{\uparrow},N_{\downarrow}}%
^{0}/E_{N_{\uparrow},N_{\downarrow}}^{0\text{,free}}.
\end{equation}
The parameter $\xi$ is defined as the thermodynamic limit for the
spin-unpolarized system,%
\begin{equation}
\xi=\lim_{N\rightarrow\infty}\xi_{N,N}.
\end{equation}

\section{Results for $\xi$ and the need for precision benchmarks}

Several experiments have measured $\xi$ using the expansion rate of $^{6}$Li
and $^{40}$K released from a harmonic trap as well as sound propagation.
\ Some recent measured values for $\xi$ are $0.32_{-10}^{+13}$
\cite{Bartenstein:2004}, $0.36(15)$ \cite{Bourdel:2004a}, $0.51(4)$
\cite{Kinast:2005}, $0.46(5)$ \cite{Partridge:2005a}, $0.46_{-12}^{+05}$
\cite{Stewart:2006}, $0.435(15)$ \cite{Joseph:2006a}, $0.41(15)$
\cite{Tarruell:2007a}, $0.41(2)$ \cite{Luo:2008a}, and $0.39(2)$
\cite{Luo:2008a}. \ A new\ preliminary measurement finds a value $0.36(1)$
\cite{Zwierlein:2011a}.

There are numerous analytical calculations of $\xi$ using a variety of
techniques such as saddle point and variational approximations
\cite{Engelbrecht:1997,Haussmann:2007a}, Pad\'{e} approximations and truncated
series methods \cite{Baker:1999dg,Heiselberg:1999,Hu:2007a}, mean field theory
with pairing \cite{Perali:2004,Kohler:2010dx}, density functional theory
extrapolated from small systems \cite{Papenbrock:2005}, renormalization group
flow \cite{Krippa:2007a}, dimensional expansions
\cite{Steele:2000qt,Schafer:2005kg,Nishida:2006a,Nishida:2006b,Chen:2006a,Arnold:2007,Nishida:2009a}%
, large-$N$ expansions \cite{Nikolic:2007}, and other methods
\cite{JChen:2006}. \ The values for $\xi$ range from $0.2$ to $0.6$ with most
predictions in the range from $0.3$ to $0.4$.

There are also many numerical calculations for $\xi$. \ The earliest
fixed-node diffusion Monte Carlo simulations for $N$ spin-up and $N$ spin-down
fermions in a periodic cube found $\xi_{N,N}$ to be $0.44(1)$ for $5\leq
N\leq21$ \cite{Carlson:2003z} and $0.42(1)$ for larger $N$
\cite{Astrakharchik:2004,Carlson:2005xy}. \ A restricted path integral Monte
Carlo calculation found similar results \cite{Akkineni:2006A}, and a
sign-restricted mean field lattice calculation yields $0.449(9)$
\cite{Juillet:2007a}. \ Another fixed-node diffusion Monte Carlo calculation
sets an upper bound for $\xi_{N,N}$ at $0.4244(1)$ for $N=33$ and $0.4339(1)$
for $N=64$ \cite{Morris:2010a}. \ A more recent fixed-node calculation sets an
upper bound for $\xi_{N,N}$ at $0.383(1)$ for $N$ between $2$ and $65$
\cite{Forbes:2010a}. \ This study includes an extrapolation to the zero-range
limit and an analysis of shell effects using density functional theory. \ We
note that methods such as fixed-node diffusion Monte Carlo provide only an
upper bound for the ground state energy. \ An unbiased estimate for the ground
state energy requires releasing the nodal constraint over a propagation time
comparable to the diffusion time for neighboring particles to cross paths.

There have also been a number of lattice simulations of two-component fermions
in the unitarity limit. \ Several lattice simulations for the average energy
at nonzero temperature have been extrapolated to the zero temperature limit.
\ The extrapolated zero temperature results from \cite{Lee:2005is,Lee:2005it}
established a bound, $0.07\leq\xi\leq0.42$. \ The results of
Ref.~\cite{Bulgac:2005a} as well as Ref.~\cite{Burovski:2006a,Burovski:2006b}
produce a value for $\xi$ in the $0.3$ to $0.5$ range. \ More recent lattice
calculations extrapolated to zero temperature yield values $\xi=0.292(24)$
\cite{Abe:2007fe,Abe:2007ff} and $\xi=0.37(5)$ \cite{Bulgac:2008c}.

In Ref.~\cite{Lee:2005fk}\ the ground state energy was calculated on the
lattice using auxiliary-field Monte Carlo and Euclidean time projection
starting from an initial state. \ The value of $\xi_{N,N}$ for $N=3,5,7,9,11$
were calculated at lattice volumes $4^{3},5^{3},6^{3}$ in units of lattice
spacing. \ From these small volumes it was estimated that $\xi=0.25(3)$. \ In
Ref.~\cite{Lee:2008xs} this lattice calculation was improved using bounded
continuous auxiliary fields. \ This calculation included an extrapolation to
the continuum limit for $\xi_{5,5}$ and $\xi_{7,7}$ using lattice volumes
$4^{3},5^{3},6^{3},$ $7^{3},8^{3}$. \ The results obtained were $\xi
_{5,5}=0.292(12)$ and $\xi_{7,7}=0.329(5)$. \ Another technique called the
symmetric heavy-light ansatz found similar values for $\xi_{N,N}$. \ While
this approach is not an \textit{ab initio} method, the agreement with the
values for $\xi_{5,5}$ and $\xi_{7,7}$ in Ref.~\cite{Lee:2008xs} were within
an error of $0.015$. \ This method gives an estimate of $\xi=0.31(1)$ in the
continuum and thermodynamic limits \cite{Lee:2007a}. \ Another extrapolation
of the same data using density functional theory to include shell effects
yields a value $\xi=0.322(2)$ \cite{Forbes:2010a}. \ Some newer but
preliminary lattice calculations using different projection and sampling
methods produce a value $\xi_{N,N}=0.412(4)$ for $N$ in the range from $8$ to
$19$ \cite{Lee:2010qp,Endres:2010sq}.

The physics of the unitarity limit is universal and can be observed in many
different systems and calculated using many different methods. \ However the
spread in experimental, analytical, and numerical evaluations for $\xi_{N,N}$
and $\xi$ highlights the need for precision benchmarks and a more careful
understanding of residual errors. \ Benchmarks at unitarity have been a
subject of much discussion at several recent workshops and programs at the
Institute for Nuclear Theory in Seattle. \ In this paper we discuss benchmarks
for four unpolarized particles in a periodic cube. \ We focus on first
principles numerical calculations for $\xi_{2,2}$ where all stochastic,
extrapolation, and systematic errors can be reliably estimated. \ The three
calculations we compare are the Hamiltonian lattice formalism using iterated
eigenvector methods, Euclidean lattice formalism with auxiliary-field
projection Monte Carlo, and continuum diffusion Monte Carlo with fixed and
released nodes.

\section{Notation and definitions}

$\ $Let $E_{N_{\uparrow},N_{\downarrow}}^{0,\text{free}}$ be the ground state
energy for $N_{\uparrow}$ up-spin and $N_{\downarrow}$ down-spin free fermions
with equal masses in a periodic cube. $\ $We write $E_{N_{\uparrow
},N_{\downarrow}}^{0}$ for the ground state energy at unitarity for the same
particle numbers, $N_{\uparrow}$ and $N_{\downarrow}$, and the same periodic
cube. \ In the introduction we defined the energy ratio,
\begin{equation}
\xi_{N_{\uparrow},N_{\downarrow}}=E_{N_{\uparrow},N_{\downarrow}}%
^{0}/E_{N_{\uparrow},N_{\downarrow}}^{0\text{,free}}. \label{few-body xi}%
\end{equation}
We should point out that there are actually two different conventions for
$\xi_{N_{\uparrow},N_{\downarrow}}$ used in the literature. \ We refer to
Eq.~(\ref{few-body xi}) as the few-body definition for the energy ratio
$\xi_{N_{\uparrow},N_{\downarrow}}$. \ This is the definition we use for all
calculations presented here.

The alternative definition for the energy ratio $\xi_{N_{\uparrow
},N_{\downarrow}}$ is what we call the thermodynamical definition. \ This
involves replacing $E_{N_{\uparrow},N_{\downarrow}}^{0,\text{free}}$ by the
formula one gets in the thermodynamic limit. \ We define the Fermi momenta and
energies in terms of the particle density,%
\begin{equation}
k_{F,\uparrow}=\left(  6\pi^{2}\frac{N_{\uparrow}}{L^{3}}\right)  ^{1/3},\quad
k_{F,\downarrow}=\left(  6\pi^{2}\frac{N_{\downarrow}}{L^{3}}\right)  ^{1/3},
\end{equation}%
\begin{equation}
E_{F,\uparrow}=\frac{k_{F,\uparrow}^{2}}{2m},\quad E_{F,\downarrow}%
=\frac{k_{F,\downarrow}^{2}}{2m}.
\end{equation}
In the thermodynamic limit the ground state energy of the non-interacting
system is%
\begin{equation}
\frac{3}{5}N_{\uparrow}E_{F,\uparrow}+\frac{3}{5}N_{\downarrow}E_{F,\downarrow
}.
\end{equation}
We use this to define the thermodynamical definition of the energy ratio,%
\begin{equation}
\xi_{N_{\uparrow},N_{\downarrow}}^{\text{thermo}}=\frac{E_{N_{\uparrow
},N_{\downarrow}}^{0}}{\frac{3}{5}N_{\uparrow}E_{F,\uparrow}+\frac{3}%
{5}N_{\downarrow}E_{F,\downarrow}}. \label{thermo_xi}%
\end{equation}

For finite $N_{\uparrow}$ and $N_{\downarrow}$ the few-body ratio
$\xi_{N_{\uparrow},N_{\downarrow}}$ and thermodynamical ratio $\xi
_{N_{\uparrow},N_{\downarrow}}^{\text{thermo}}$ differ due to shell effects in
the non-interacting system. \ There are several calculations in the literature
using each of these two alternative definitions. \ In
Table~\ref{xi_conversion} we have tabulated the conversion between the two
definitions for several values of particle number with $N_{\uparrow
}=N_{\downarrow}$.

\begin{table}[tb]
\caption{Conversion factor between the two ground state ratios $\xi
_{N_{\uparrow},N_{\downarrow}}$ and $\xi_{N_{\uparrow},N_{\downarrow}%
}^{\text{thermo}}$ for various values $N_{\uparrow}=N_{\downarrow}$.}%
\label{xi_conversion}
$%
\begin{tabular}
[c]{|c|c|c|c|c|}\cline{1-2}\cline{4-5}%
$N_{\uparrow}=N_{\downarrow}$ & $\xi_{N_{\uparrow},N_{\downarrow}}%
/\xi_{N_{\uparrow},N_{\downarrow}}^{\text{thermo}}$ & \  & $N_{\uparrow
}=N_{\downarrow}$ & $\xi_{N_{\uparrow},N_{\downarrow}}/\xi_{N_{\uparrow
},N_{\downarrow}}^{\text{thermo}}$\\\cline{1-2}\cline{4-5}%
$2$ & $0.7331$ & \  & $8$ & $0.9236$\\\cline{1-2}\cline{4-5}%
$3$ & $0.7204$ & \  & $9$ & $0.8991$\\\cline{1-2}\cline{4-5}%
$4$ & $0.7758$ & \  & $10$ & $0.8931$\\\cline{1-2}\cline{4-5}%
$5$ & $0.8439$ & \  & $16$ & $0.9774$\\\cline{1-2}\cline{4-5}%
$6$ & $0.9149$ & \  & $24$ & $1.0246$\\\cline{1-2}\cline{4-5}%
$7$ & $0.9858$ & \  & $32$ & $1.0064$\\\cline{1-2}\cline{4-5}%
\end{tabular}
\ \ \ $\end{table}

\section{Hamiltonian lattice with sparse-matrix eigenvector iteration}

\subsection{Formalism and notation}

Let $\vec{n}$ denote spatial lattice points on a three-dimensional $L\times
L\times L$ periodic cube. $\ $We use lattice units where physical quantities
are multiplied by powers of the spatial lattice spacing to make the
combination dimensionless. \ The two-component fermions are labelled as
spin-up and spin-down, the lattice annihilation operators are written as
$a_{\uparrow}(\vec{n})$ and $a_{\downarrow}(\vec{n})$. \ We start with the
free non-relativistic lattice Hamiltonian,
\begin{equation}
H_{\text{free}}=\frac{3}{m}\sum_{\vec{n},i=\uparrow,\downarrow}a_{i}^{\dagger
}(\vec{n})a_{i}(\vec{n})-\frac{1}{2m}\sum_{l=1,2,3}\sum_{\vec{n}%
,i=\uparrow,\downarrow}\left[  a_{i}^{\dagger}(\vec{n})a_{i}(\vec{n}+\hat
{l})+a_{i}^{\dagger}(\vec{n})a_{i}(\vec{n}-\hat{l})\right]  . \label{Hfree}%
\end{equation}
We define the spin-density operators%
\begin{equation}
\rho_{\uparrow}(\vec{n})=a_{\uparrow}^{\dagger}(\vec{n})a_{\uparrow}(\vec{n}),
\label{rho_up}%
\end{equation}%
\begin{equation}
\rho_{\downarrow}(\vec{n})=a_{\downarrow}^{\dagger}(\vec{n})a_{\downarrow
}(\vec{n}). \label{rho_down}%
\end{equation}
We consider two different lattice Hamiltonians each of which yield the
unitarity limit in the low-energy limit. \ The first Hamiltonian, $H_{1}$, has
a single-site contact interaction,
\begin{equation}
H_{1}=H_{\text{free}}+C_{1}\sum_{\vec{n}}\rho_{\uparrow}(\vec{n}%
)\rho_{\downarrow}(\vec{n}). \label{H1_latt}%
\end{equation}
The coefficient of $C_{1}$ is tuned to set the S-wave scattering length
$a_{0}$ to infinity. \ The second Hamiltonian, $H_{2}$, has a contact
interaction as well as nearest-neighbor interaction terms,
\begin{align}
H_{2}  &  =H_{\text{free}}+C_{2}\sum_{\vec{n}}\rho_{\uparrow}(\vec{n}%
)\rho_{\downarrow}(\vec{n})\nonumber\\
&  +C_{2}^{\prime}\sum_{l=1,2,3}\sum_{\vec{n}}\left[  \rho_{\uparrow}(\vec
{n})\rho_{\downarrow}(\vec{n}+\hat{l})+\rho_{\uparrow}(\vec{n}+\hat{l}%
)\rho_{\downarrow}(\vec{n})\right]  . \label{H2_latt}%
\end{align}
The coefficients $C_{2}$ and $C_{2}^{\prime}$ are tuned so that $a_{0}$ goes
to infinity while the S-wave effective range parameter $r_{0}$ vanishes.

\bigskip We use L\"{u}scher's formula
\cite{Luscher:1986pf,Beane:2003da,Seki:2005ns,Borasoy:2006qn} to determine the
unknown interaction coefficients $C_{1}$, $C_{2}$, and $C_{2}^{\prime}$.
\ L\"{u}scher's formula relates the two-particle energy levels in a length $L$
periodic cube to the S-wave phase shift,%
\begin{equation}
p\cot\delta_{0}(p)=\frac{1}{\pi L}S\left(  \eta\right)  ,\qquad\eta=\left(
\frac{Lp}{2\pi}\right)  ^{2},
\end{equation}
where $S(\eta)$ is the three-dimensional zeta function,%
\begin{equation}
S(\eta)=\lim_{\Lambda\rightarrow\infty}\left[  \sum_{\vec{n}}\frac
{\theta(\Lambda^{2}-\vec{n}^{2})}{\vec{n}^{2}-\eta}-4\pi\Lambda\right]  .
\end{equation}
In terms of $\eta$, the energy of the two-particle scattering state is%
\begin{equation}
E_{\text{pole}}=\frac{p^{2}}{m}=\frac{\eta}{m}\left(  \frac{2\pi}{L}\right)
^{2}.
\end{equation}

\begin{figure}[ptb]%
\centering
\includegraphics[
height=2.527in,
width=2.8271in
]%
{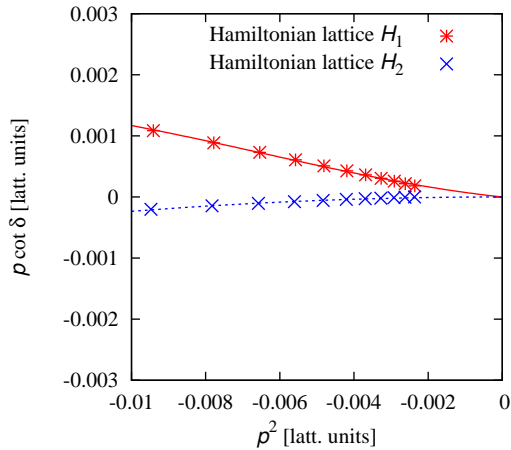}%
\caption{(Color online) Plot of $p\cot\delta_{0}(p)$ versus $p^{2}$ for the
lattice Hamiltonians $H_{1}$ and $H_{2}$.}%
\label{pcot_hamiltonian}%
\end{figure}

The S-wave effective range expansion gives%
\begin{equation}
p\cot\delta_{0}(p)=-\frac{1}{a_{0}}+\frac{1}{2}r_{0}p^{2}+O(p^{4}).
\end{equation}
Setting $a_{0}$ to infinity requires $p\cot\delta_{0}(p)$ to vanish at
threshold. \ Setting both $a_{0}$ to infinity and $r_{0}$ to zero requires
that $p\cot\delta_{0}(p)$ is $O(p^{4})$ near threshold. \ The plots for
$p\cot\delta_{0}(p)$ versus $p^{2}$ are shown in Fig.~\ref{pcot_hamiltonian}.
\ The values we find for the interaction coefficients are%
\begin{equation}
mC_{1}=-3.9570,
\end{equation}%
\begin{equation}
mC_{2}=-3.7235,\qquad mC_{2}^{\prime}=-0.3008.
\end{equation}
%

\subsection{Results for the four-particle benchmark}

\begin{figure}[ptb]%
\centering
\includegraphics[
height=3.0294in,
width=2.8262in
]%
{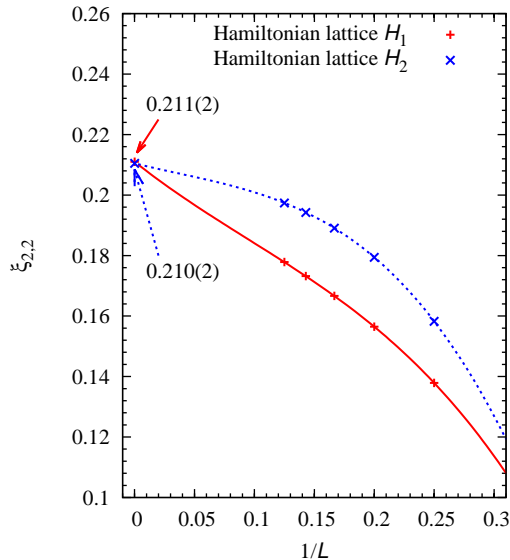}%
\caption{(Color online) Ground state energy ratio $\xi_{2,2}$ for lattice
Hamiltonians $H_{1}$ and $H_{2}$. \ We show results for values $L=4,5,6,7,8$,
and extrapolate to the infinite volume limit.}%
\label{xi22_hamiltonian}%
\end{figure}

Using the Lanczos algorithm for sparse-matrix eigenvector iteration
\cite{Lanczos:1950}, we have computed the ground state energy for two spin-up
and two spin-down particles in a periodic cube of length $L$. \ For both
lattice Hamiltonians, $H_{1}$ and $H_{2}$, we have computed $\xi_{2,2}$ as
defined in Eq.~(\ref{few-body xi}) for values $L=4,5,6,7,8$. \ The results are
shown in Fig.~(\ref{xi22_hamiltonian}).%

We have fitted the data using polynomials in $1/L$ up to third order and
extrapolate to the infinite $L$ limit with an estimated extrapolation error of
$\pm0.002$. \ We note that this extrapolation should remove all measurable
lattice discretization effects. \ For $H_{1}$ we find%
\begin{equation}
\xi_{2,2}=0.211(2), \label{H1}%
\end{equation}
and for $H_{2}$ we get%
\begin{equation}
\xi_{2,2}=0.210(2). \label{H2}%
\end{equation}
The agreement between these two independent calculations is consistent with
our estimate of the systematic errors.\ 

The third-degree polynomial extrapolation is made possible by the high
precision data obtained for each $L$ using Lanczos eigenvector iteration.
\ For the Monte Carlo data appearing later in our discussion we use only
linear extrapolations in $1/L$. \ For the $H_{2}$ data we note the small slope
in $1/L$ near $1/L=0$. \ This is expected due to the effective range $r_{0}$
being set to zero for $H_{2}$. \ The small amount of linear dependence in
$1/L$ that remains is likely due to other lattice artifacts such as the
breaking of Galilean invariance \cite{Lee:2007jd}.

\section{Euclidean lattice with auxiliary-field projection Monte Carlo}

\subsection{Formalism and notation}

For the Euclidean lattice calculation we use the normal-ordered
transfer-matrix formalism used in
Ref.~\cite{Lee:2005fk,Lee:2006hr,Lee:2008fa,Lee:2008xs}. \ Normal-ordering
refers to the rearrangement of operators with annihilation operators on the
right and creation operators on the left. \ This prescription is useful in
that it provides an exact relation between Grassmann path integration and the
operator formalism \cite{Creutz:1988wv,Creutz:1999zy}. \ The details of the
application of this correspondence can be found in
Ref.~\cite{Lee:2006hr,Lee:2008fa,Lee:2008xs}. \ As before we use lattice units
which correspond with multiplying physical quantities by the corresponding
power of the spatial lattice spacing to make the combination dimensionless.
\ We write $\alpha_{t}=a_{t}$/$a$ for the dimensionless ratio of the temporal
lattice spacing to spatial lattice spacing. \ For fermion mass $m$, we take
the ratio of lattice spacings so that $m^{-1}\alpha_{t}=0.1109$.

\begin{figure}[ptb]%
\centering
\includegraphics[
height=2.527in,
width=2.8271in
]%
{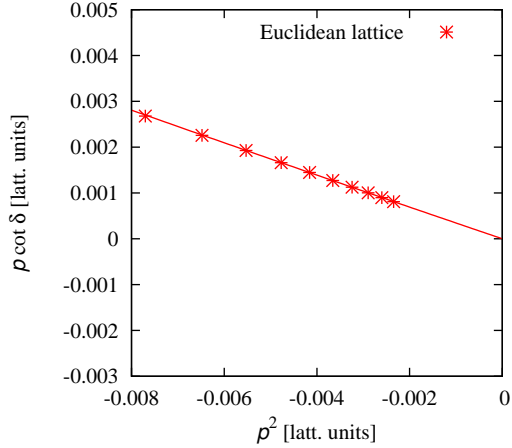}%
\caption{(Color online) Plot of $p\cot\delta_{0}(p)$ versus $p^{2}$ for the
Euclidean-time lattice formalism.}%
\label{pcot_compact}%
\end{figure}

We start with the normal-ordered transfer matrix operator,%
\begin{equation}
M=\;:\exp\left[  -H_{\text{free}}\alpha_{t}-C\alpha_{t}\sum_{\vec{n}}%
\rho_{\uparrow}(\vec{n})\rho_{\downarrow}(\vec{n})\right]  :,
\label{transfer_noaux}%
\end{equation}
The free lattice Hamiltonian $H_{\text{free}}$ was defined in Eq.~(\ref{Hfree}%
) and the spin densities were defined Eq.~(\ref{rho_up},\ref{rho_down}).
\ Just as in the Hamiltonian lattice calculation we use L\"{u}scher's formula
to determine the unknown coefficient $C$. \ Setting the S-wave scattering
length to infinity, we find
\begin{equation}
mC=-3.4938.
\end{equation}
In Fig.~(\ref{pcot_compact}) we plot $p\cot\delta_{0}(p)$ versus $p^{2}$ at
unitarity. \

For the Monte Carlo simulations we use the bounded continuous auxiliary-field
transformation introduced in Ref.~\cite{Lee:2008xs}. \ This transformation was
shown in Ref.~\cite{Lee:2008xs} to have performance advantages over other
continuous and discrete auxiliary-field transformations. \ We can write the
transfer matrix as%
\begin{equation}
M=\prod\limits_{\vec{n}}\left[  \frac{1}{2\pi}\int_{-\pi}^{+\pi}ds(\vec
{n},n_{t})\right]  M(s,n_{t})
\end{equation}
where%
\begin{equation}
M(s,n_{t})=\,\colon\exp\left\{  -H_{\text{free}}\alpha_{t}+\sum_{\vec{n}}%
\sqrt{-2C\alpha_{t}}\sin\left[  s(\vec{n},n_{t})\right]  \cdot\left[
\rho_{\uparrow}(\vec{n})+\rho_{\downarrow}(\vec{n})\right]  \right\}  \colon.
\label{transfer_auxiliary}%
\end{equation}

Let $\left\vert \Psi_{2,2}^{\text{init}}\right\rangle $ a Slater-determinant
of single-particle normal modes composed of two spin-up fermions and two
spin-down fermions,%
\begin{equation}
\left\vert \Psi_{2,2}^{\text{init}}\right\rangle =\left\vert \psi
_{1}\right\rangle \wedge\left\vert \psi_{2}\right\rangle \wedge\left\vert
\psi_{3}\right\rangle \wedge\left\vert \psi_{4}\right\rangle .
\end{equation}
For the calculations presented here we choose%
\begin{equation}
\left\vert \psi_{1}\right\rangle =\sqrt{\frac{1}{L^{3}}}\sum_{\vec{n}%
}a_{\uparrow}^{\dagger}(\vec{n})\left\vert 0\right\rangle ,\quad\left\vert
\psi_{2}\right\rangle =\sqrt{\frac{2}{L^{3}}}\sum_{\vec{n}}\cos(2\pi
n_{3}/L)a_{\uparrow}^{\dagger}(\vec{n})\left\vert 0\right\rangle ,
\end{equation}%
\begin{equation}
\left\vert \psi_{3}\right\rangle =\sqrt{\frac{1}{L^{3}}}\sum_{\vec{n}%
}a_{\downarrow}^{\dagger}(\vec{n})\left\vert 0\right\rangle ,\quad\left\vert
\psi_{4}\right\rangle =\sqrt{\frac{2}{L^{3}}}\sum_{\vec{n}}\cos(2\pi
n_{3}/L)a_{\downarrow}^{\dagger}(\vec{n})\left\vert 0\right\rangle .
\end{equation}
$\left\vert \psi_{1}\right\rangle $ and $\left\vert \psi_{3}\right\rangle $
are constant-valued throughout the periodic box, while $\left\vert \psi
_{2}\right\rangle $ and $\left\vert \psi_{4}\right\rangle $ are standing waves
with wavelength $L$ along the $z$-axis. \ We construct the Euclidean-time
projection amplitude%
\begin{equation}
Z_{2,2}(t)\equiv\prod\limits_{\vec{n},n_{t}}\left[  \frac{1}{2\pi}\int_{-\pi
}^{+\pi}ds(\vec{n},n_{t})\right]  \left\langle \Psi_{2,2}^{\text{init}%
}\right\vert M(s,L_{t}-1)\cdot\cdots\cdot M(s,0)\left\vert \Psi_{2,2}%
^{\text{init}}\right\rangle ,
\end{equation}
where the Euclidean projection time $t$ equals $L_{t}$ times the temporal
lattice spacing.

Each $M(s,n_{t})$ consists of only single-particle operators interacting with
the background auxiliary field. \ Therefore we find%
\begin{equation}
\left\langle \Psi_{2,2}^{\text{init}}\right\vert M(s,L_{t}-1)\cdot\cdots\cdot
M(s,0)\left\vert \Psi_{2,2}^{\text{init}}\right\rangle =\left[  \det
\mathbf{M}(s,t)\right]  ^{2}, \label{detsquare}%
\end{equation}
where%
\begin{equation}
\left[  \mathbf{M}(s,t)\right]  _{k^{\prime}k}=\left\langle \psi_{k^{\prime}%
}\right\vert M(s,L_{t}-1)\cdot\cdots\cdot M(s,0)\left\vert \psi_{k}%
\right\rangle ,
\end{equation}
for matrix indices $k,k^{\prime}=1,2,3,4$. \ We define a $t$-dependent energy
expectation value,%
\begin{equation}
E_{2,2}(t)=\frac{1}{\alpha_{t}}\ln\frac{Z_{2,2}(t-\alpha_{t})}{Z_{2,2}(t)}.
\label{E_nn_t}%
\end{equation}
We can also express the $t$-dependent expectation value as a fraction of the
non-interacting ground state energy,%
\begin{equation}
\xi_{2,2}(t)=E_{2,2}(t)/E_{2,2}^{0,\text{free}}.
\end{equation}
The ground state energy $E_{N,N}^{0}$ is given by the limit%
\begin{equation}
E_{2,2}^{0}=\lim_{t\rightarrow\infty}E_{2,2}(t), \label{ground_state}%
\end{equation}
and the desired few-body energy ratio can be computed as the limit
\begin{equation}
\xi_{2,2}=\lim_{t\rightarrow\infty}\xi_{2,2}(t).
\end{equation}

\subsection{Results for the four-particle benchmark}

\begin{figure}[ptb]%
\centering
\includegraphics[
height=2.527in,
width=2.8271in
]%
{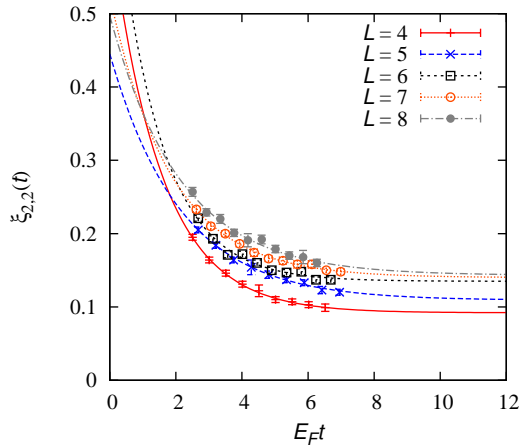}%
\caption{(Color online) The lattice data for $\xi_{2,2}(t)$ versus $E_{F}t$
for $L=4,5,6,7,8$. \ Also shown are the results of the asymptotic fits.}%
\label{xi_compact_lt}%
\end{figure}

For the calculation of $\xi_{2,2}(t)$ we use the lattice dimensions
$L^{3}\times L_{t}$ shown in Table~\ref{dimensions_22}. \ \begin{table}[tb]
\caption{Lattice dimensions $L^{3}\times L_{t}$ used in calculations for
$\xi_{2,2}(t).$}%
\label{dimensions_22}
\begin{tabular}
[c]{|c|c|c|c|c|c|}\hline
$L^{3}$ & $4^{3}$ & $5^{3}$ & $6^{3}$ & $7^{3}$ & $8^{3}$\\\hline
$L_{t}$ &
\begin{tabular}
[c]{c}%
$30$\\
$36$\\
$\vdots$\\
$78$%
\end{tabular}
&
\begin{tabular}
[c]{c}%
$50$\\
$60$\\
$\vdots$\\
$130$%
\end{tabular}
&
\begin{tabular}
[c]{c}%
$72$\\
$84$\\
$\vdots$\\
$180$%
\end{tabular}
&
\begin{tabular}
[c]{c}%
$96$\\
$112$\\
$\vdots$\\
$256$%
\end{tabular}
&
\begin{tabular}
[c]{c}%
$120$\\
$140$\\
$\vdots$\\
$300$%
\end{tabular}
\\\hline
\end{tabular}
\end{table}The simulations are run with $2048$ processors each independently
generating $3000$ hybrid Monte Carlo trajectories
\cite{Scalettar:1986uy,Gottlieb:1987mq,Duane:1987de}. \ To extract $\xi_{2,2}$
we perform a least-squares fit of $\xi_{2,2}(t)$ to the asymptotic form,%
\begin{equation}
\xi_{2,2}(t)=\xi_{2,2}+be^{-\delta E\,t}.
\end{equation}
This exponential form takes into account the contribution from higher-energy
states and is the same method used in
Ref.~\cite{Lee:2005fk,Lee:2006hr,Lee:2008fa,Lee:2008xs}. \ We focus on
measuring the ground state energy accurately and ignore the numerically small
contributions hidden in the far asymptotic tail of $\xi_{2,2}(t)$. \ We
determine $b$, $\delta E$, and $\xi_{2,2}$ from least-squares fitting over the
range $E_{F}t=2$ to $E_{F}t=6$. \ The lattice data for $\xi_{2,2}(t)$ together
with the asymptotic fits are shown in Fig.~\ref{xi_compact_lt}.%

\begin{figure}[ptb]%
\centering
\includegraphics[
height=2.527in,
width=2.8271in
]%
{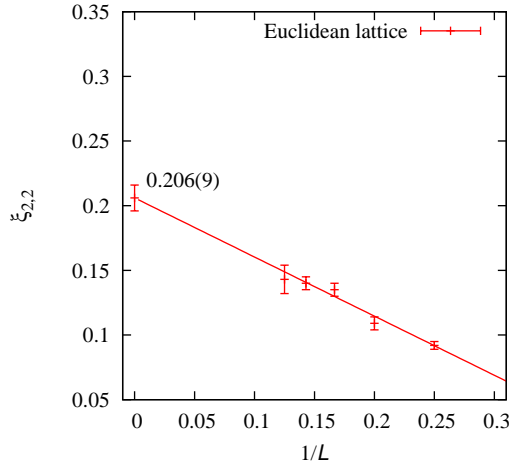}%
\caption{(Color online) Results for $\xi_{2,2}$ for $L=4,5,6,7,8$ plotted
versus $L^{-1}$. \ The lattice results are extrapolated to the continuum limit
$L\rightarrow\infty.$}%
\label{xi22_compact}%
\end{figure}

We use the lattice results for $\xi_{2,2}$ with $L=4,5,6,7,8$ to extrapolate
to the continuum limit $L\rightarrow\infty$. \ In Fig.~\ref{xi22_compact} we
show the lattice results for $\xi_{2,2}$ for $L=4,5,6,7,8$ plotted versus
$L^{-1}$. \ We expect a dependence on $L$ arising from effects such as the
effective range correction and lattice cutoff effects. \ Using a linear
extrapolation in $L^{-1}$, we obtain the continuum limit value%
\begin{equation}
\xi_{2,2}=0.206(9). \label{Euclidean}%
\end{equation}
This result using Euclidean lattice projection Monte Carlo is in agreement
with the Hamiltonian lattice results in Eq.~(\ref{H1},\ref{H2}).%

\section{Diffusion Monte Carlo with fixed and released nodes}

\subsection{Formalism and notation}

We now discuss diffusion Monte Carlo calculations for the same benchmark
system of four particles at unitarity in a periodic cube with length $L$.
\ This time, however, we use continuous variables and consider the evolution
as a function of Euclidean time as a diffusion Monte Carlo process using an
ensemble of random walkers. \ An introduction to the basic techniques can be
found in Ref.~\cite{Mitas:2001}. \ For the interaction between spin-up and
spin-down fermions we use a P\"{o}schl-Teller potential tuned to infinite
S-wave scattering length. \ For fermion mass $m$, the form of the potential is%
\begin{equation}
V(r)=-\frac{2}{m}\frac{\mu^{2}}{\cosh^{2}(\mu r)},
\end{equation}
where the momentum scale $\mu$ determines the S-wave effective range
parameter,%
\begin{equation}
r_{0}=2\mu^{-1}.
\end{equation}
The unitarity limit corresponds with taking the limit $\mu\rightarrow\infty$.
\ For the unpolarized four-particle system we let $\mathbf{R}$ be the set of
individual particle coordinates,
\begin{equation}
\mathbf{R}=\left\{  \vec{r}_{1_{\uparrow}},\vec{r}_{1_{\downarrow}},\vec
{r}_{2_{\uparrow}},\vec{r}_{2_{\downarrow}}\right\}  .
\end{equation}
We use a BCS-type pairing wavefunction projected onto two spin-up and two
spin-down fermions. \ The wavefunction $\Psi_{\text{BCS}}\mathbf{(R)}$ can be
written as a $2\times2$ Slater determinant,
\begin{equation}
\Psi_{\text{BCS}}\mathbf{(R)}=~\det\left[
\begin{array}
[c]{cc}%
\phi(\vec{r}_{1_{\uparrow}}-\vec{r}_{1_{\downarrow}}) & \phi(\vec
{r}_{1_{\uparrow}}-\vec{r}_{2_{\downarrow}})\\
\phi(\vec{r}_{2_{\uparrow}}-\vec{r}_{1_{\downarrow}}) & \phi(\vec
{r}_{2_{\uparrow}}-\vec{r}_{2_{\downarrow}})
\end{array}
\right]  ,
\end{equation}
where $\phi(\vec{r})$ is the pairing function. \ The trial wavefunction,
$\Psi_{T}(\mathbf{R)}$, is given as a product of $\Psi_{\text{BCS}%
}\mathbf{(R)}$ times a Jastrow factor \cite{Jastrow:1955a},
\begin{equation}
\Psi_{T}(\mathbf{R)}=\Psi_{\text{BCS}}\mathbf{(R})\exp\left[  J(\mathbf{R)}%
\right]  \mathbf{.}%
\end{equation}
The Jastrow factor incorporates particle correlations and is useful in
reducing stochastic errors in the Monte Carlo calculation
\cite{Umrigar:1988a,Schmidt:1990a,Mitas:1994a,Grossman:1995b}. \ We use a
product of Gaussian functions for the Jastrow factor. \ The exponents of these
Gaussian functions are tuned to minimize the combination of the variational
energy and the variance of the local energy in the variational Monte Carlo.
\ We note that the positive definite function $\exp\left[  J(\mathbf{R)}%
\right]  $ has no effect on the nodal structure of $\Psi_{T}(\mathbf{R}).$

To determine the pairing function $\phi(\vec{r})$ we use an approach similar
to that in Ref.~\cite{Carlson:2003z}. \ We use an ansatz which is a
superposition of Gaussian functions with periodic copies,%
\begin{equation}
\phi(\vec{r})=\sum_{k}d_{k}\sum_{s_{x},s_{y},s_{z}=-1}^{1}e^{-\frac{\alpha
_{k}}{2}(x+s_{x}L)^{2}}e^{-\frac{\alpha_{k}}{2}(y+s_{y}L)^{2}}e^{-\frac
{\alpha_{k}}{2}(z+s_{z}L)^{2}}.
\end{equation}
Here $\vec{r}=(x,y,z)$, and $d_{k}$ and $\alpha_{k}$ are variational
parameters. \ Gaussian functions from the nearest periodic images at distance
$L$ make small contributions while periodic copies further away can be
neglected. \ This construction has the advantage of providing more flexibility
for pairing orbitals in the box while keeping the orbitals smooth over the
periodic boundary with zero derivative. \ The Jastrow factor $\exp
[J(\mathbf{R})]$ is constructed similarly. \ The calculation is performed
using standard variational Monte Carlo sampling of the square of the trial
function \cite{Mitas:2001,Bajdich:2009}.

For the diffusion Monte Carlo calculation we use a large ensemble of
forward-propagating random walkers with population branching processes
determined by the local energy and guided by the trial wavefunction $\Psi
_{T}(\mathbf{R)}$. \ The local density of random walkers gives a statistical
estimate of the product of the propagated quantum wavefunction $\Psi
(\mathbf{R})$ times the $\Psi_{T}(\mathbf{R)}$ trial function. \ In the
fixed-node diffusion Monte Carlo (FN-DMC) calculation, the nodal structure of
$\Psi(\mathbf{R})$ is fixed by $\Psi_{T}(\mathbf{R)}$, and the product
$\Psi(\mathbf{R})$ $\Psi_{T}(\mathbf{R)}$ is therefore positive definite.
Since the trial function is known explicitly and analytically it is then
possible to extract the energetics of the lowest fermionic state within the
fixed-node boundary conditions \cite{Mitas:2001,Bajdich:2009}.

The fixed-node calculation sets an upper bound on the ground state energy.
\ To measure the quality of the upper bound we release the nodal constraints
\cite{Ceperley:1984}. \ For the released-node diffusion Monte Carlo
calculation (RN-DMC) we use a positive-definite guiding profile for the
diffusion of random walkers. \ In the calculations presented here we consider
a one-parameter family of guiding profiles,
\begin{equation}
\Psi_{G}^{\alpha}(\mathbf{R})=\sqrt{\Psi_{T}^{2}(\mathbf{R})+\alpha
\left\langle \Psi_{T}^{2}\right\rangle }.
\end{equation}
The dimensionless parameter $\alpha$ controls the rate of diffusion across the
nodal boundaries of $\Psi_{T}(\mathbf{R})$, and $\left\langle \Psi_{T}%
^{2}\right\rangle $ is the average value of $\Psi_{T}^{2}(\mathbf{R_{0}})$
evaluated over all $\mathbf{R_{0}}$, where $\mathbf{R_{0}}$ is the
configuration right after the nodal release process.

\subsection{Results for the four-particle benchmark}

\begin{figure}[ptb]%
\centering
\includegraphics[
height=3.0251in,
width=3.3468in
]%
{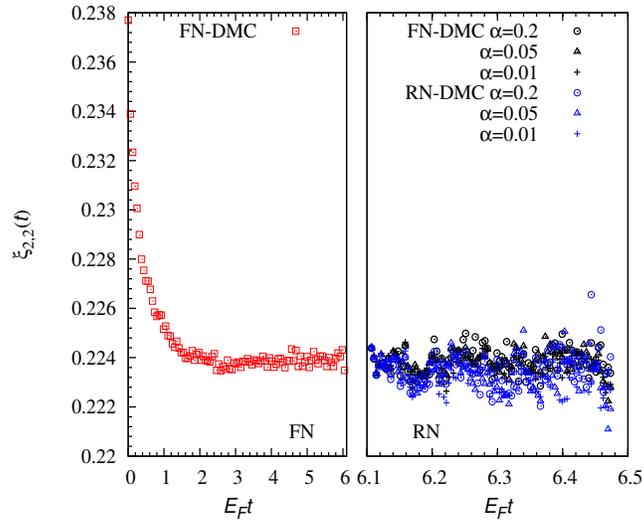}%
\caption{(Color online) Fixed-node and released-node diffusion\ Monte Carlo
results for $r_{0}/L=0.025$ and $\alpha=0.01$, $0.05$, $0.2.$}%
\label{released_node}%
\end{figure}

In Fig.~\ref{released_node} we show fixed-node (FN-DMC) and released-node
(RN-DMC) diffusion\ Monte Carlo results for two spin-up and spin-down fermions
in a periodic cube at unitarity. \ The ratio of the effective range parameter
to the length of the cube is $r_{0}/L=1/40$. \ We use parameters $\alpha
=0.01$, $0.05$, $0.2$ for the guiding profile $\Psi_{G}^{\alpha}(\mathbf{R})$.
\ With each nodal crossing the associated weights of the random walkers change
sign, leading to a sign cancellation problem which grows exponentially with
Euclidean propagation time $t$. \ For the calculations presented here we have
measured the released-node correction starting from the fixed-node result up
to propagation time of $t=0.4E_{F}^{-1}$. \ As seen in
Fig.~\ref{released_node}, the decrease in $\xi_{2,2}$ is less than $0.002$
over the duration of the released-node time propagation.%

\begin{figure}[ptb]%
\centering
\includegraphics[
height=3.026in,
width=3.173in
]%
{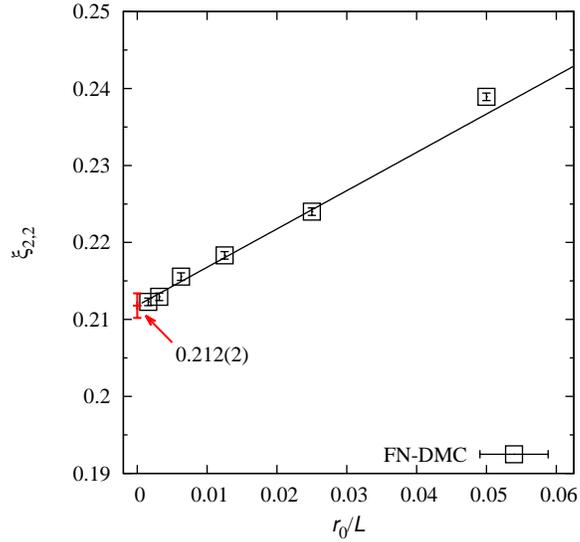}%
\caption{(Color online) Fixed-node diffusion Monte Carlo results for
$\xi_{2,2}$ versus the $r_{0}/L$. \ The data is extrapolated to the limit of
zero effective range.}%
\label{xi22_diffusion}%
\end{figure}

We have repeated the fixed-node and released-node calculations of $\xi_{2,2}$
for values $r_{0}/L=1/20$, $1/40$, $1/80$, $1/160$, $1/320$, $1/640$. \ The
fixed-node results for $\xi_{2,2}$ versus the $r_{0}/L$ are shown in
Fig.~\ref{xi22_diffusion}. \ Using a linear fit in $r_{0}/L$, we extrapolated
to the limit of zero effective range.%

In the zero-range limit we get the final result%
\begin{equation}
\xi_{2,2}=0.212(2) \label{fixed-node result}%
\end{equation}
in the fixed-node approximation. \ The error estimate includes the stochastic
error, zero-range extrapolation error, and errors due to time step
discretization and other small residual effects. \ The released-node
calculations are quantitatively similar to the results shown in
Fig.~\ref{released_node} for $r_{0}/L=1/40$. \ The decrease in $\xi_{2,2}$ is
less than $0.002$ over a propagation time of $0.4E_{F}^{-1}$.\ \ We note that
$E_{F}^{-1}$ is the characteristic time scale required for two neighboring
particles of the same spin to cross paths. \ Given the stochastic noise in the
released-node calculation results, it is difficult to pin down a correction to
the fixed-node result from the nodal release. \ However a reasonable
conservative estimate is that the upper bound set by the fixed-node
calculation is less than $0.002/0.4=0.005$ above the actual value. \ This
appears confirmed by the agreement of Eq.~(\ref{fixed-node result}) with the
values for $\xi_{2,2}$ obtain using Hamiltonian lattice eigenvector iteration
and Euclidean lattice Monte Carlo.

\section{Summary and discussion}

We have presented benchmark calculations for four unpolarized particles in a
periodic cube using three different methods. \ In the Hamiltonian lattice
formalism with iterated eigenvector methods, we obtained%
\begin{equation}
\xi_{2,2}=0.211(2)
\end{equation}
using the Hamiltonian $H_{1}$ defined in Eq.~(\ref{H1_latt}), and%
\begin{equation}
\xi_{2,2}=0.210(2)
\end{equation}
using the Hamiltonian $H_{2}$ defined in Eq.~(\ref{H2_latt}). \ Using the
Euclidean lattice formalism with auxiliary-field projection Monte Carlo we
found the result
\begin{equation}
\xi_{2,2}=0.206(9).
\end{equation}
With fixed-node diffusion Monte Carlo in the continuum we extracted the upper
bound%
\begin{equation}
\xi_{2,2}\leq0.212(2).
\end{equation}
The release-node Monte Carlo calculation shows a decrease in $\xi_{2,2}$ that
is less than $0.002$ over a propagation time of $0.4E_{F}^{-1}$. \ We estimate
that the upper bound set by the fixed-node calculation is less than $0.005$
above the actual value. \ All three methods agree within estimated errors.
\ The unpolarized four-particle benchmarks presented here should be useful for
testing and calibrating residual errors for other numerical methods and
perhaps also analytical calculations. \ We note that the comparison requires
using the few-body definition of $\xi_{2,2}$ in Eq.~(\ref{few-body xi}).
$\ $If the thermodynamical definition in Eq.~(\ref{thermo_xi}) is used then
the conversion factor is $\xi_{2,2}=0.7331\xi_{2,2}^{\text{thermo}}$.

We note the importance of continuum limit extrapolations for lattice
calculations and zero-range limit extrapolations for continuum diffusion Monte
Carlo calculations. \ The importance of continuum limit extrapolations in
lattice calculations was already noted and measured in Ref.~\cite{Lee:2008xs}.
\ In the fixed-node diffusion Monte Carlo calculations presented here we find
that $r_{0}/L$ needs to be less than $0.03$ in order to obtain a value for
$\xi_{2,2}$ accurate to a relative error of $0.1$. \ This is consistent with
the range dependence found in Ref.~\cite{Forbes:2010a}. \ After discussion
with the authors of Ref.~\cite{Forbes:2010a}, we are informed that they
obtained an upper bound for $\xi_{2,2}$ from fixed-node diffusion Monte Carlo
agreeing within two significant digits with the results reported here.

In this paper the four-particle benchmark was chosen to allow comparisons
among several very different numerical methods. \ For larger $N=N_{\uparrow
}=N_{\downarrow}$ systems it is unfortunately not possible to use iterated
eigenvector methods due to exponential $L^{6N-3}$ scaling in memory. \ However
the Euclidean lattice Monte Carlo and diffusion Monte Carlo methods extend
readily to larger systems, and it is useful to comment on the relationship
between the four-particle calculations discussed here and calculations in
larger systems at unitarity.

The four-particle results presented here should be useful for comparisons of
different lattice Monte Carlo calculations using different lattice actions and
algorithms. \ We note that there is a significant correction produced by the
extrapolation to the continuum limit. \ For some lattice actions this
extrapolation decreases the ground state energy while for others it increases
the ground state energy. \ In all cases it is important that stochastic and
systematic errors are sufficiently small for each chosen lattice spacing so
that the continuum extrapolation can be done accurately. \ The same lattice
methods used here to find $\xi_{2,2}=0.206(9)$ were also used to determine
$\xi_{5,5}=0.292(12)$ and $\xi_{7,7}=0.329(5)$ in Ref.~\cite{Lee:2008xs}. \ In
order to benchmark different lattice Monte Carlo methods, one starting point
would be to test agreement with each of these values for $\xi_{N,N}$.

After completion of the original manuscript of this paper, the fixed-node
diffusion Monte Carlo methods presented here were also applied to larger
$N=N_{\uparrow}=N_{\downarrow}$ systems at unitarity. \ In
Ref.~\cite{Li:2011b} the results $\xi_{7,7}\leq0.407(2)$, $\xi_{19,19}%
\leq0.409(3),$ and $\xi_{33,33}\leq0.398(3)$ are presented. \ These results
are comparable to the upper bounds found in Ref.~\cite{Forbes:2010a}, and in
each case a significant reduction in the ground state energy is seen when
extrapolating to the zero range limit. \ For the released-node calculations in
these larger systems, the exponential severity of sign cancellations makes it
difficult to extract data for Euclidean propagation time $t$ greater than
$N^{-1}E_{F}^{-1}$. \ As noted above, $E_{F}^{-1}$ is the characteristic time
scale required for two neighboring particles of the same spin to cross paths.
\ For $N=2$ we extrapolated to get a bound on the decrease in $\xi_{2,2}$ over
a propagation time of $E_{F}^{-1}$. \ For larger $N$ the extrapolation from
propagation time $N^{-1}E_{F}^{-1}$ to time $E_{F}^{-1}$ cannot be done
reliably, and this is seen already for the case $N=7$ \cite{Li:2011b}.

We note that there remains a significant gap between the lattice result
$\xi_{7,7}=0.329(5)$ and the fixed-node upper bound $\xi_{7,7}\leq0.407(2)$.
\ This discrepancy must be better understood. \ A good starting point would be
to make benchmark comparisons for other small systems, $N=3,4,5$. \ First it
should be established that different lattice calculations agree on the values
for $\xi_{3,3},$ $\xi_{4,4},$ and $\xi_{5,5}$. \ Next the difference between
lattice and fixed-node results should be measured for each $N$. \ The key
question then is if this difference can be resolved by released-node
calculations for smaller values of $N$. \ Given the agreement for $\xi_{2,2}$,
there is some reason to suggest that this may be possible.

\section*{Acknowledgements}

We thank Michael Forbes, Stefano Gandolfi, Alex Gezerlis, and Hans-Werner
Hammer for useful discussions. \ We acknowledge the 2010 Institute for Nuclear
Theory program on Simulations and Symmetries: Cold Atoms, QCD, and Few-hadron
Systems organized by Daniel Phillips, Hans-Werner Hammer, and Martin Savage.
\ Some discussions during that program provided motivation for this
collaborative work. \ Partial financial support from the Deutsche
Forschungsgemeinschaft (SFB/TR 16), Helmholtz Association (contract number
VH-VI-231), BMBF\ (grant 06BN9006), DOE LANL subcontract 81279-001-10, NSF
grants DMR-0804549 and OCI-0904794, ARO, and U.S. Department of Energy
(DE-FG02-03ER41260) are acknowledged. \ This work was further supported by the
EU HadronPhysics2 project \textquotedblleft Study of strongly interacting
matter\textquotedblright. \ The computational resources for this project were
provided by the J\"{u}lich Supercomputing Centre at the Forschungszentrum
J\"{u}lich, by ORNL under INCITE program, by NSF TRAC allocation at TACC, and
by the Center for High Performance Computing at NC\ State University.

\bibliographystyle{apsrev}
\bibliography{References_fourparticle}

\begin{thebibliography}{77}
\expandafter\ifx\csname natexlab\endcsname\relax\def\natexlab#1{#1}\fi
\expandafter\ifx\csname bibnamefont\endcsname\relax
  \def\bibnamefont#1{#1}\fi
\expandafter\ifx\csname bibfnamefont\endcsname\relax
  \def\bibfnamefont#1{#1}\fi
\expandafter\ifx\csname citenamefont\endcsname\relax
  \def\citenamefont#1{#1}\fi
\expandafter\ifx\csname url\endcsname\relax
  \def\url#1{\texttt{#1}}\fi
\expandafter\ifx\csname urlprefix\endcsname\relax\def\urlprefix{URL }\fi
\providecommand{\bibinfo}[2]{#2}
\providecommand{\eprint}[2][]{\url{#2}}

\bibitem[{\citenamefont{Eagles}(1969)}]{Eagles:1969PR}
\bibinfo{author}{\bibfnamefont{D.~M.} \bibnamefont{Eagles}},
  \bibinfo{journal}{Phys. Rev.} \textbf{\bibinfo{volume}{186}},
  \bibinfo{pages}{456} (\bibinfo{year}{1969}).

\bibitem[{\citenamefont{Leggett}(1980)}]{Leggett:1980pro}
\bibinfo{author}{\bibfnamefont{A.~J.} \bibnamefont{Leggett}}, in
  \emph{\bibinfo{booktitle}{Modern Trends in the Theory of Condensed Matter.
  Proceedings of the XVIth Karpacz Winter School of Theoretical Physics,
  Karpacz, Poland, 1980}} (\bibinfo{publisher}{Springer-Verlag, Berlin},
  \bibinfo{year}{1980}), p.~\bibinfo{pages}{13}.

\bibitem[{\citenamefont{Nozieres and Schmitt-Rink}(1985)}]{Nozieres:1985JLTP}
\bibinfo{author}{\bibfnamefont{P.}~\bibnamefont{Nozieres}} \bibnamefont{and}
  \bibinfo{author}{\bibfnamefont{S.}~\bibnamefont{Schmitt-Rink}},
  \bibinfo{journal}{J. Low Temp. Phys.} \textbf{\bibinfo{volume}{59}},
  \bibinfo{pages}{195} (\bibinfo{year}{1985}).

\bibitem[{\citenamefont{O'Hara et~al.}(2002)\citenamefont{O'Hara, Hemmer, Gehm,
  Granade, and Thomas}}]{O'Hara:2002}
\bibinfo{author}{\bibfnamefont{K.~M.} \bibnamefont{O'Hara}},
  \bibinfo{author}{\bibfnamefont{S.~L.} \bibnamefont{Hemmer}},
  \bibinfo{author}{\bibfnamefont{M.~E.} \bibnamefont{Gehm}},
  \bibinfo{author}{\bibfnamefont{S.~R.} \bibnamefont{Granade}},
  \bibnamefont{and} \bibinfo{author}{\bibfnamefont{J.~E.}
  \bibnamefont{Thomas}}, \bibinfo{journal}{Science}
  \textbf{\bibinfo{volume}{298}}, \bibinfo{pages}{2179} (\bibinfo{year}{2002}).

\bibitem[{\citenamefont{Gupta et~al.}(2003)\citenamefont{Gupta, Hadzibabic,
  Zwierlein, Stan, Dieckmann, Schunck, van Kempen, Verhaar, and
  Ketterle}}]{Gupta:2002}
\bibinfo{author}{\bibfnamefont{S.}~\bibnamefont{Gupta}},
  \bibinfo{author}{\bibfnamefont{Z.}~\bibnamefont{Hadzibabic}},
  \bibinfo{author}{\bibfnamefont{M.~W.} \bibnamefont{Zwierlein}},
  \bibinfo{author}{\bibfnamefont{C.~A.} \bibnamefont{Stan}},
  \bibinfo{author}{\bibfnamefont{K.}~\bibnamefont{Dieckmann}},
  \bibinfo{author}{\bibfnamefont{C.~H.} \bibnamefont{Schunck}},
  \bibinfo{author}{\bibfnamefont{E.~G.~M.} \bibnamefont{van Kempen}},
  \bibinfo{author}{\bibfnamefont{B.~J.} \bibnamefont{Verhaar}},
  \bibnamefont{and} \bibinfo{author}{\bibfnamefont{W.}~\bibnamefont{Ketterle}},
  \bibinfo{journal}{Science} \textbf{\bibinfo{volume}{300}},
  \bibinfo{pages}{1723} (\bibinfo{year}{2003}).

\bibitem[{\citenamefont{Regal and Jin}(2003)}]{Regal:2003}
\bibinfo{author}{\bibfnamefont{C.~A.} \bibnamefont{Regal}} \bibnamefont{and}
  \bibinfo{author}{\bibfnamefont{D.~S.} \bibnamefont{Jin}},
  \bibinfo{journal}{Phys. Rev. Lett.} \textbf{\bibinfo{volume}{90}},
  \bibinfo{pages}{230404} (\bibinfo{year}{2003}).

\bibitem[{\citenamefont{Bourdel et~al.}(2003)\citenamefont{Bourdel, Cubizolles,
  Khaykovich, Magalhaes, Kokkelmans, Shlyapnikov, and Salomon}}]{Bourdel:2003}
\bibinfo{author}{\bibfnamefont{T.}~\bibnamefont{Bourdel}},
  \bibinfo{author}{\bibfnamefont{J.}~\bibnamefont{Cubizolles}},
  \bibinfo{author}{\bibfnamefont{L.}~\bibnamefont{Khaykovich}},
  \bibinfo{author}{\bibfnamefont{K.~M.~F.} \bibnamefont{Magalhaes}},
  \bibinfo{author}{\bibfnamefont{S.~J. J. M.~F.} \bibnamefont{Kokkelmans}},
  \bibinfo{author}{\bibfnamefont{G.~V.} \bibnamefont{Shlyapnikov}},
  \bibnamefont{and} \bibinfo{author}{\bibfnamefont{C.}~\bibnamefont{Salomon}},
  \bibinfo{journal}{Phys. Rev. Lett.} \textbf{\bibinfo{volume}{91}},
  \bibinfo{pages}{020402} (\bibinfo{year}{2003}).

\bibitem[{\citenamefont{Gehm et~al.}(2003)\citenamefont{Gehm, Hemmer, Granade,
  O'Hara, and Thomas}}]{Gehm:2003}
\bibinfo{author}{\bibfnamefont{M.~E.} \bibnamefont{Gehm}},
  \bibinfo{author}{\bibfnamefont{S.~L.} \bibnamefont{Hemmer}},
  \bibinfo{author}{\bibfnamefont{S.~R.} \bibnamefont{Granade}},
  \bibinfo{author}{\bibfnamefont{K.~M.} \bibnamefont{O'Hara}},
  \bibnamefont{and} \bibinfo{author}{\bibfnamefont{J.~E.}
  \bibnamefont{Thomas}}, \bibinfo{journal}{Phys. Rev.}
  \textbf{\bibinfo{volume}{A68}}, \bibinfo{pages}{011401(R)}
  (\bibinfo{year}{2003}).

\bibitem[{\citenamefont{Bartenstein et~al.}(2004)\citenamefont{Bartenstein,
  Altmeyer, Riedl, Jochim, Chin, Hecker~Denschlag, and
  Grimm}}]{Bartenstein:2004}
\bibinfo{author}{\bibfnamefont{M.}~\bibnamefont{Bartenstein}},
  \bibinfo{author}{\bibfnamefont{A.}~\bibnamefont{Altmeyer}},
  \bibinfo{author}{\bibfnamefont{S.}~\bibnamefont{Riedl}},
  \bibinfo{author}{\bibfnamefont{S.}~\bibnamefont{Jochim}},
  \bibinfo{author}{\bibfnamefont{C.}~\bibnamefont{Chin}},
  \bibinfo{author}{\bibfnamefont{J.}~\bibnamefont{Hecker~Denschlag}},
  \bibnamefont{and} \bibinfo{author}{\bibfnamefont{R.}~\bibnamefont{Grimm}},
  \bibinfo{journal}{Phys. Rev. Lett.} \textbf{\bibinfo{volume}{92}},
  \bibinfo{pages}{120401} (\bibinfo{year}{2004}), \eprint{cond-mat/0401109v2}.

\bibitem[{\citenamefont{Bourdel et~al.}(2004)\citenamefont{Bourdel, Khaykovich,
  Cubizolles, Zhang, Chevy, Teichmann, Tarruell, Kokkelmans, and
  Salomon}}]{Bourdel:2004a}
\bibinfo{author}{\bibfnamefont{T.}~\bibnamefont{Bourdel}},
  \bibinfo{author}{\bibfnamefont{L.}~\bibnamefont{Khaykovich}},
  \bibinfo{author}{\bibfnamefont{J.}~\bibnamefont{Cubizolles}},
  \bibinfo{author}{\bibfnamefont{J.}~\bibnamefont{Zhang}},
  \bibinfo{author}{\bibfnamefont{F.}~\bibnamefont{Chevy}},
  \bibinfo{author}{\bibfnamefont{M.}~\bibnamefont{Teichmann}},
  \bibinfo{author}{\bibfnamefont{L.}~\bibnamefont{Tarruell}},
  \bibinfo{author}{\bibfnamefont{S.~J. J. M.~F.} \bibnamefont{Kokkelmans}},
  \bibnamefont{and} \bibinfo{author}{\bibfnamefont{C.}~\bibnamefont{Salomon}},
  \bibinfo{journal}{Phys. Rev. Lett.} \textbf{\bibinfo{volume}{93}},
  \bibinfo{pages}{050401} (\bibinfo{year}{2004}), \eprint{cond-mat/0403091v3}.

\bibitem[{\citenamefont{Kinast et~al.}(2005)\citenamefont{Kinast, Turlapov,
  Thomas, Chen, Stajic, and Levin}}]{Kinast:2005}
\bibinfo{author}{\bibfnamefont{J.}~\bibnamefont{Kinast}},
  \bibinfo{author}{\bibfnamefont{A.}~\bibnamefont{Turlapov}},
  \bibinfo{author}{\bibfnamefont{J.~E.} \bibnamefont{Thomas}},
  \bibinfo{author}{\bibfnamefont{Q.}~\bibnamefont{Chen}},
  \bibinfo{author}{\bibfnamefont{J.}~\bibnamefont{Stajic}}, \bibnamefont{and}
  \bibinfo{author}{\bibfnamefont{K.}~\bibnamefont{Levin}},
  \bibinfo{journal}{Science} \textbf{\bibinfo{volume}{307}},
  \bibinfo{pages}{1296} (\bibinfo{year}{2005}), \eprint{cond-mat/0502087}.

\bibitem[{\citenamefont{{Partridge} et~al.}(2006)\citenamefont{{Partridge},
  {Li}, {Kamar}, {Liao}, and {Hulet}}}]{Partridge:2005a}
\bibinfo{author}{\bibfnamefont{G.~B.} \bibnamefont{{Partridge}}},
  \bibinfo{author}{\bibfnamefont{W.}~\bibnamefont{{Li}}},
  \bibinfo{author}{\bibfnamefont{R.~I.} \bibnamefont{{Kamar}}},
  \bibinfo{author}{\bibfnamefont{Y.}~\bibnamefont{{Liao}}}, \bibnamefont{and}
  \bibinfo{author}{\bibfnamefont{R.~G.} \bibnamefont{{Hulet}}},
  \bibinfo{journal}{Science} \textbf{\bibinfo{volume}{311}},
  \bibinfo{pages}{503} (\bibinfo{year}{2006}), \eprint{arXiv:cond-mat/0511752}.

\bibitem[{\citenamefont{Stewart et~al.}(2006)\citenamefont{Stewart, Gaebler,
  Regal, and Jin}}]{Stewart:2006}
\bibinfo{author}{\bibfnamefont{J.~T.} \bibnamefont{Stewart}},
  \bibinfo{author}{\bibfnamefont{J.~P.} \bibnamefont{Gaebler}},
  \bibinfo{author}{\bibfnamefont{C.~A.} \bibnamefont{Regal}}, \bibnamefont{and}
  \bibinfo{author}{\bibfnamefont{D.~S.} \bibnamefont{Jin}},
  \bibinfo{journal}{Phys. Rev. Lett.} \textbf{\bibinfo{volume}{97}},
  \bibinfo{pages}{220406} (\bibinfo{year}{2006}), \eprint{cond-mat/0607776}.

\bibitem[{\citenamefont{Joseph et~al.}(2007)\citenamefont{Joseph, Clancy, Luo,
  Kinast, Turlapov, and Thomas}}]{Joseph:2006a}
\bibinfo{author}{\bibfnamefont{J.}~\bibnamefont{Joseph}},
  \bibinfo{author}{\bibfnamefont{B.}~\bibnamefont{Clancy}},
  \bibinfo{author}{\bibfnamefont{L.}~\bibnamefont{Luo}},
  \bibinfo{author}{\bibfnamefont{J.}~\bibnamefont{Kinast}},
  \bibinfo{author}{\bibfnamefont{A.}~\bibnamefont{Turlapov}}, \bibnamefont{and}
  \bibinfo{author}{\bibfnamefont{J.~E.} \bibnamefont{Thomas}},
  \bibinfo{journal}{Phys. Rev. Lett.} \textbf{\bibinfo{volume}{98}},
  \bibinfo{pages}{170401} (\bibinfo{year}{2007}), \eprint{cond-mat/0612567v1}.

\bibitem[{\citenamefont{Tarruell et~al.}(2008)\citenamefont{Tarruell,
  Teichmann, Mckeever, Bourdel, Cubizolles, Khaykovich, Zhang, Navon, Chevy,
  and Salomon}}]{Tarruell:2007a}
\bibinfo{author}{\bibfnamefont{L.}~\bibnamefont{Tarruell}},
  \bibinfo{author}{\bibfnamefont{M.}~\bibnamefont{Teichmann}},
  \bibinfo{author}{\bibfnamefont{J.}~\bibnamefont{Mckeever}},
  \bibinfo{author}{\bibfnamefont{T.}~\bibnamefont{Bourdel}},
  \bibinfo{author}{\bibfnamefont{J.}~\bibnamefont{Cubizolles}},
  \bibinfo{author}{\bibfnamefont{L.}~\bibnamefont{Khaykovich}},
  \bibinfo{author}{\bibfnamefont{J.}~\bibnamefont{Zhang}},
  \bibinfo{author}{\bibfnamefont{N.}~\bibnamefont{Navon}},
  \bibinfo{author}{\bibfnamefont{F.}~\bibnamefont{Chevy}}, \bibnamefont{and}
  \bibinfo{author}{\bibfnamefont{C.}~\bibnamefont{Salomon}}, in
  \emph{\bibinfo{booktitle}{Ultra-Cold Fermi Gases: Proceedings of the
  International School of Physics "Enrico Fermi", Course CLXIV}}
  (\bibinfo{publisher}{I O S Press, Incorporated}, \bibinfo{year}{2008}),
  \eprint{cond-mat/0701181}.

\bibitem[{\citenamefont{Luo and Thomas}(2009)}]{Luo:2008a}
\bibinfo{author}{\bibfnamefont{L.}~\bibnamefont{Luo}} \bibnamefont{and}
  \bibinfo{author}{\bibfnamefont{J.~E.} \bibnamefont{Thomas}},
  \bibinfo{journal}{Journal of Low Temperature Physics}
  \textbf{\bibinfo{volume}{154}}, \bibinfo{pages}{1} (\bibinfo{year}{2009}),
  \eprint{arXiv:0811.1159 [cond-mat.other]}.

\bibitem[{\citenamefont{Zwierlein}(2011)}]{Zwierlein:2011a}
\bibinfo{author}{\bibfnamefont{M.}~\bibnamefont{Zwierlein}}
  (\bibinfo{year}{2011}), \bibinfo{note}{presented at March APS Meeting,
  Dallas, 2011}.

\bibitem[{\citenamefont{Engelbrecht et~al.}(1997)\citenamefont{Engelbrecht,
  Randeria, and de~Melo}}]{Engelbrecht:1997}
\bibinfo{author}{\bibfnamefont{J.~R.} \bibnamefont{Engelbrecht}},
  \bibinfo{author}{\bibfnamefont{M.}~\bibnamefont{Randeria}}, \bibnamefont{and}
  \bibinfo{author}{\bibfnamefont{C.~S.} \bibnamefont{de~Melo}},
  \bibinfo{journal}{Phys. Rev.} \textbf{\bibinfo{volume}{B55}},
  \bibinfo{pages}{15153} (\bibinfo{year}{1997}).

\bibitem[{\citenamefont{Haussmann et~al.}(2007)\citenamefont{Haussmann,
  Rantner, Cerrito, and Zwerger}}]{Haussmann:2007a}
\bibinfo{author}{\bibfnamefont{R.}~\bibnamefont{Haussmann}},
  \bibinfo{author}{\bibfnamefont{W.}~\bibnamefont{Rantner}},
  \bibinfo{author}{\bibfnamefont{S.}~\bibnamefont{Cerrito}}, \bibnamefont{and}
  \bibinfo{author}{\bibfnamefont{W.}~\bibnamefont{Zwerger}},
  \bibinfo{journal}{Phys. Rev. A} \textbf{\bibinfo{volume}{75}},
  \bibinfo{pages}{023610} (\bibinfo{year}{2007}).

\bibitem[{\citenamefont{Baker}(1999)}]{Baker:1999dg}
\bibinfo{author}{\bibfnamefont{G.~A.} \bibnamefont{Baker}},
  \bibinfo{journal}{Phys. Rev.} \textbf{\bibinfo{volume}{C60}},
  \bibinfo{pages}{054311} (\bibinfo{year}{1999}).

\bibitem[{\citenamefont{Heiselberg}(2001)}]{Heiselberg:1999}
\bibinfo{author}{\bibfnamefont{H.}~\bibnamefont{Heiselberg}},
  \bibinfo{journal}{Phys. Rev.} \textbf{\bibinfo{volume}{A63}},
  \bibinfo{pages}{043606} (\bibinfo{year}{2001}), \eprint{cond-mat/0002056}.

\bibitem[{\citenamefont{{Hu} et~al.}(2007)\citenamefont{{Hu}, {Drummond}, and
  {Liu}}}]{Hu:2007a}
\bibinfo{author}{\bibfnamefont{H.}~\bibnamefont{{Hu}}},
  \bibinfo{author}{\bibfnamefont{P.~D.} \bibnamefont{{Drummond}}},
  \bibnamefont{and} \bibinfo{author}{\bibfnamefont{X.}~\bibnamefont{{Liu}}},
  \bibinfo{journal}{Nature Physics} \textbf{\bibinfo{volume}{3}},
  \bibinfo{pages}{469} (\bibinfo{year}{2007}), \eprint{arXiv:cond-mat/0701744}.

\bibitem[{\citenamefont{Perali et~al.}(2004)\citenamefont{Perali, Pieri, and
  Strinati}}]{Perali:2004}
\bibinfo{author}{\bibfnamefont{A.}~\bibnamefont{Perali}},
  \bibinfo{author}{\bibfnamefont{P.}~\bibnamefont{Pieri}}, \bibnamefont{and}
  \bibinfo{author}{\bibfnamefont{G.~C.} \bibnamefont{Strinati}},
  \bibinfo{journal}{Phys. Rev. Lett.} \textbf{\bibinfo{volume}{93}},
  \bibinfo{pages}{100404} (\bibinfo{year}{2004}).

\bibitem[{\citenamefont{Kohler}(2010)}]{Kohler:2010dx}
\bibinfo{author}{\bibfnamefont{H.~S.} \bibnamefont{Kohler}}
  (\bibinfo{year}{2010}), \eprint{1008.3884}.

\bibitem[{\citenamefont{Papenbrock}(2005)}]{Papenbrock:2005}
\bibinfo{author}{\bibfnamefont{T.}~\bibnamefont{Papenbrock}},
  \bibinfo{journal}{Phys. Rev.} \textbf{\bibinfo{volume}{A72}},
  \bibinfo{pages}{041603(R)} (\bibinfo{year}{2005}), \eprint{cond-mat/0507183}.

\bibitem[{\citenamefont{Krippa}(2009)}]{Krippa:2007a}
\bibinfo{author}{\bibfnamefont{B.}~\bibnamefont{Krippa}}, \bibinfo{journal}{J.
  Phys.} \textbf{\bibinfo{volume}{A42}}, \bibinfo{pages}{465002}
  (\bibinfo{year}{2009}), \eprint{arXiv:0704.3984v4 [cond-mat.supr-con]}.

\bibitem[{\citenamefont{Steele}(2000)}]{Steele:2000qt}
\bibinfo{author}{\bibfnamefont{J.~V.} \bibnamefont{Steele}}
  (\bibinfo{year}{2000}), \eprint{nucl-th/0010066}.

\bibitem[{\citenamefont{Sch{\"a}fer et~al.}(2005)\citenamefont{Sch{\"a}fer,
  Kao, and Cotanch}}]{Schafer:2005kg}
\bibinfo{author}{\bibfnamefont{T.}~\bibnamefont{Sch{\"a}fer}},
  \bibinfo{author}{\bibfnamefont{C.-W.} \bibnamefont{Kao}}, \bibnamefont{and}
  \bibinfo{author}{\bibfnamefont{S.~R.} \bibnamefont{Cotanch}},
  \bibinfo{journal}{Nucl. Phys.} \textbf{\bibinfo{volume}{A762}},
  \bibinfo{pages}{82} (\bibinfo{year}{2005}), \eprint{nucl-th/0504088}.

\bibitem[{\citenamefont{Nishida and Son}(2006)}]{Nishida:2006a}
\bibinfo{author}{\bibfnamefont{Y.}~\bibnamefont{Nishida}} \bibnamefont{and}
  \bibinfo{author}{\bibfnamefont{D.~T.} \bibnamefont{Son}},
  \bibinfo{journal}{Phys. Rev. Lett.} \textbf{\bibinfo{volume}{97}},
  \bibinfo{pages}{050403} (\bibinfo{year}{2006}), \eprint{cond-mat/0604500}.

\bibitem[{\citenamefont{Nishida and Son}(2007)}]{Nishida:2006b}
\bibinfo{author}{\bibfnamefont{Y.}~\bibnamefont{Nishida}} \bibnamefont{and}
  \bibinfo{author}{\bibfnamefont{D.~T.} \bibnamefont{Son}},
  \bibinfo{journal}{Phys. Rev.} \textbf{\bibinfo{volume}{A75}},
  \bibinfo{pages}{063617} (\bibinfo{year}{2007}), \eprint{cond-mat/0607835}.

\bibitem[{\citenamefont{Chen and Nakano}(2007)}]{Chen:2006a}
\bibinfo{author}{\bibfnamefont{J.-W.} \bibnamefont{Chen}} \bibnamefont{and}
  \bibinfo{author}{\bibfnamefont{E.}~\bibnamefont{Nakano}},
  \bibinfo{journal}{Phys. Rev.} \textbf{\bibinfo{volume}{A75}},
  \bibinfo{pages}{043620} (\bibinfo{year}{2007}), \eprint{cond-mat/0610011}.

\bibitem[{\citenamefont{Arnold et~al.}(2007)\citenamefont{Arnold, Drut, and
  Son}}]{Arnold:2007}
\bibinfo{author}{\bibfnamefont{P.}~\bibnamefont{Arnold}},
  \bibinfo{author}{\bibfnamefont{J.~E.} \bibnamefont{Drut}}, \bibnamefont{and}
  \bibinfo{author}{\bibfnamefont{D.~T.} \bibnamefont{Son}},
  \bibinfo{journal}{Phys. Rev.} \textbf{\bibinfo{volume}{A75}},
  \bibinfo{pages}{043605} (\bibinfo{year}{2007}), \eprint{cond-mat/0608477}.

\bibitem[{\citenamefont{Nishida}(2009)}]{Nishida:2009a}
\bibinfo{author}{\bibfnamefont{Y.}~\bibnamefont{Nishida}},
  \bibinfo{journal}{Phys. Rev. A} \textbf{\bibinfo{volume}{79}},
  \bibinfo{pages}{013627} (\bibinfo{year}{2009}).

\bibitem[{\citenamefont{Nikolic and Sachdev}(2007)}]{Nikolic:2007}
\bibinfo{author}{\bibfnamefont{P.}~\bibnamefont{Nikolic}} \bibnamefont{and}
  \bibinfo{author}{\bibfnamefont{S.}~\bibnamefont{Sachdev}},
  \bibinfo{journal}{Phys. Rev.} \textbf{\bibinfo{volume}{A75}},
  \bibinfo{pages}{033608} (\bibinfo{year}{2007}), \eprint{cond-mat/0609106}.

\bibitem[{\citenamefont{Chen}(2007)}]{JChen:2006}
\bibinfo{author}{\bibfnamefont{J.}~\bibnamefont{Chen}},
  \bibinfo{journal}{Chinese Phys. Lett.} \textbf{\bibinfo{volume}{24}},
  \bibinfo{pages}{1825} (\bibinfo{year}{2007}), \eprint{nucl-th/0602065}.

\bibitem[{\citenamefont{Carlson et~al.}(2003)\citenamefont{Carlson, Chang,
  Pandharipande, and Schmidt}}]{Carlson:2003z}
\bibinfo{author}{\bibfnamefont{J.}~\bibnamefont{Carlson}},
  \bibinfo{author}{\bibfnamefont{S.~Y.} \bibnamefont{Chang}},
  \bibinfo{author}{\bibfnamefont{V.~R.} \bibnamefont{Pandharipande}},
  \bibnamefont{and} \bibinfo{author}{\bibfnamefont{K.}~\bibnamefont{Schmidt}},
  \bibinfo{journal}{Phys. Rev. Lett.} \textbf{\bibinfo{volume}{91}},
  \bibinfo{pages}{50401} (\bibinfo{year}{2003}), \eprint{physics/0303094}.

\bibitem[{\citenamefont{Astrakharchik et~al.}(2004)\citenamefont{Astrakharchik,
  Boronat, Casulleras, and Giorgini}}]{Astrakharchik:2004}
\bibinfo{author}{\bibfnamefont{G.~E.} \bibnamefont{Astrakharchik}},
  \bibinfo{author}{\bibfnamefont{J.}~\bibnamefont{Boronat}},
  \bibinfo{author}{\bibfnamefont{J.}~\bibnamefont{Casulleras}},
  \bibnamefont{and} \bibinfo{author}{\bibfnamefont{S.}~\bibnamefont{Giorgini}},
  \bibinfo{journal}{Phys. Rev. Lett.} \textbf{\bibinfo{volume}{93}},
  \bibinfo{pages}{200404} (\bibinfo{year}{2004}), \eprint{cond-mat/0406113}.

\bibitem[{\citenamefont{Carlson and Reddy}(2005)}]{Carlson:2005xy}
\bibinfo{author}{\bibfnamefont{J.}~\bibnamefont{Carlson}} \bibnamefont{and}
  \bibinfo{author}{\bibfnamefont{S.}~\bibnamefont{Reddy}},
  \bibinfo{journal}{Phys. Rev. Lett.} \textbf{\bibinfo{volume}{95}},
  \bibinfo{pages}{060401} (\bibinfo{year}{2005}), \eprint{cond-mat/0503256}.

\bibitem[{\citenamefont{Akkineni et~al.}(2007)\citenamefont{Akkineni, Ceperley,
  and Trivedi}}]{Akkineni:2006A}
\bibinfo{author}{\bibfnamefont{V.~K.} \bibnamefont{Akkineni}},
  \bibinfo{author}{\bibfnamefont{D.~M.} \bibnamefont{Ceperley}},
  \bibnamefont{and} \bibinfo{author}{\bibfnamefont{N.}~\bibnamefont{Trivedi}},
  \bibinfo{journal}{Phys. Rev. B} \textbf{\bibinfo{volume}{76}},
  \bibinfo{pages}{165116} (\bibinfo{year}{2007}).

\bibitem[{\citenamefont{Juillet}(2007)}]{Juillet:2007a}
\bibinfo{author}{\bibfnamefont{O.}~\bibnamefont{Juillet}},
  \bibinfo{journal}{New J. Phys.} \textbf{\bibinfo{volume}{9}},
  \bibinfo{pages}{163} (\bibinfo{year}{2007}), \eprint{cond-mat/0609063}.

\bibitem[{\citenamefont{Morris et~al.}(2010)\citenamefont{Morris,
  L\'opez~R\'\i{}os, and Needs}}]{Morris:2010a}
\bibinfo{author}{\bibfnamefont{A.~J.} \bibnamefont{Morris}},
  \bibinfo{author}{\bibfnamefont{P.}~\bibnamefont{L\'opez~R\'\i{}os}},
  \bibnamefont{and} \bibinfo{author}{\bibfnamefont{R.~J.} \bibnamefont{Needs}},
  \bibinfo{journal}{Phys. Rev. A} \textbf{\bibinfo{volume}{81}},
  \bibinfo{pages}{033619} (\bibinfo{year}{2010}).

\bibitem[{\citenamefont{{McNeil Forbes} et~al.}(2010)\citenamefont{{McNeil
  Forbes}, {Gandolfi}, and {Gezerlis}}}]{Forbes:2010a}
\bibinfo{author}{\bibfnamefont{M.}~\bibnamefont{{McNeil Forbes}}},
  \bibinfo{author}{\bibfnamefont{S.}~\bibnamefont{{Gandolfi}}},
  \bibnamefont{and}
  \bibinfo{author}{\bibfnamefont{A.}~\bibnamefont{{Gezerlis}}}
  (\bibinfo{year}{2010}), \eprint{arXiv:1011.2197 [cond-mat.quant.gas]}.

\bibitem[{\citenamefont{Lee and Sch{\"a}fer}(2006{\natexlab{a}})}]{Lee:2005is}
\bibinfo{author}{\bibfnamefont{D.}~\bibnamefont{Lee}} \bibnamefont{and}
  \bibinfo{author}{\bibfnamefont{T.}~\bibnamefont{Sch{\"a}fer}},
  \bibinfo{journal}{Phys. Rev.} \textbf{\bibinfo{volume}{C73}},
  \bibinfo{pages}{015201} (\bibinfo{year}{2006}{\natexlab{a}}),
  \eprint{nucl-th/0509017}.

\bibitem[{\citenamefont{Lee and Sch{\"a}fer}(2006{\natexlab{b}})}]{Lee:2005it}
\bibinfo{author}{\bibfnamefont{D.}~\bibnamefont{Lee}} \bibnamefont{and}
  \bibinfo{author}{\bibfnamefont{T.}~\bibnamefont{Sch{\"a}fer}},
  \bibinfo{journal}{Phys. Rev.} \textbf{\bibinfo{volume}{C73}},
  \bibinfo{pages}{015202} (\bibinfo{year}{2006}{\natexlab{b}}),
  \eprint{nucl-th/0509018}.

\bibitem[{\citenamefont{Bulgac et~al.}(2006)\citenamefont{Bulgac, Drut, and
  Magierski}}]{Bulgac:2005a}
\bibinfo{author}{\bibfnamefont{A.}~\bibnamefont{Bulgac}},
  \bibinfo{author}{\bibfnamefont{J.~E.} \bibnamefont{Drut}}, \bibnamefont{and}
  \bibinfo{author}{\bibfnamefont{P.}~\bibnamefont{Magierski}},
  \bibinfo{journal}{Phys. Rev. Lett.} \textbf{\bibinfo{volume}{96}},
  \bibinfo{pages}{090404} (\bibinfo{year}{2006}), \eprint{cond-mat/0505374}.

\bibitem[{\citenamefont{Burovski
  et~al.}(2006{\natexlab{a}})\citenamefont{Burovski, Prokofev, Svistunov, and
  Troyer}}]{Burovski:2006a}
\bibinfo{author}{\bibfnamefont{E.}~\bibnamefont{Burovski}},
  \bibinfo{author}{\bibfnamefont{N.}~\bibnamefont{Prokofev}},
  \bibinfo{author}{\bibfnamefont{B.}~\bibnamefont{Svistunov}},
  \bibnamefont{and} \bibinfo{author}{\bibfnamefont{M.}~\bibnamefont{Troyer}},
  \bibinfo{journal}{Phys. Rev. Lett.} \textbf{\bibinfo{volume}{96}},
  \bibinfo{pages}{160402} (\bibinfo{year}{2006}{\natexlab{a}}),
  \eprint{cond-mat/0602224}.

\bibitem[{\citenamefont{Burovski
  et~al.}(2006{\natexlab{b}})\citenamefont{Burovski, Prokofev, Svistunov, and
  Troyer}}]{Burovski:2006b}
\bibinfo{author}{\bibfnamefont{E.}~\bibnamefont{Burovski}},
  \bibinfo{author}{\bibfnamefont{N.}~\bibnamefont{Prokofev}},
  \bibinfo{author}{\bibfnamefont{B.}~\bibnamefont{Svistunov}},
  \bibnamefont{and} \bibinfo{author}{\bibfnamefont{M.}~\bibnamefont{Troyer}},
  \bibinfo{journal}{New J. Phys.} \textbf{\bibinfo{volume}{8}},
  \bibinfo{pages}{153} (\bibinfo{year}{2006}{\natexlab{b}}),
  \eprint{cond-mat/0605350}.

\bibitem[{\citenamefont{Abe and Seki}(2009{\natexlab{a}})}]{Abe:2007fe}
\bibinfo{author}{\bibfnamefont{T.}~\bibnamefont{Abe}} \bibnamefont{and}
  \bibinfo{author}{\bibfnamefont{R.}~\bibnamefont{Seki}},
  \bibinfo{journal}{Phys. Rev.} \textbf{\bibinfo{volume}{C79}},
  \bibinfo{pages}{054002} (\bibinfo{year}{2009}{\natexlab{a}}),
  \eprint{arXiv:0708.2523 [nucl-th]}.

\bibitem[{\citenamefont{Abe and Seki}(2009{\natexlab{b}})}]{Abe:2007ff}
\bibinfo{author}{\bibfnamefont{T.}~\bibnamefont{Abe}} \bibnamefont{and}
  \bibinfo{author}{\bibfnamefont{R.}~\bibnamefont{Seki}},
  \bibinfo{journal}{Phys. Rev.} \textbf{\bibinfo{volume}{C79}},
  \bibinfo{pages}{054003} (\bibinfo{year}{2009}{\natexlab{b}}),
  \eprint{arXiv:0708.2524 [nucl-th]}.

\bibitem[{\citenamefont{Bulgac et~al.}(2008)\citenamefont{Bulgac, Drut, and
  Magierski}}]{Bulgac:2008c}
\bibinfo{author}{\bibfnamefont{A.}~\bibnamefont{Bulgac}},
  \bibinfo{author}{\bibfnamefont{J.~E.} \bibnamefont{Drut}}, \bibnamefont{and}
  \bibinfo{author}{\bibfnamefont{P.}~\bibnamefont{Magierski}},
  \bibinfo{journal}{Phys. Rev. A} \textbf{\bibinfo{volume}{78}},
  \bibinfo{pages}{023625} (\bibinfo{year}{2008}), \eprint{arXiv:0803.3238
  [cond-mat.stat-mech]}.

\bibitem[{\citenamefont{Lee}(2006)}]{Lee:2005fk}
\bibinfo{author}{\bibfnamefont{D.}~\bibnamefont{Lee}}, \bibinfo{journal}{Phys.
  Rev.} \textbf{\bibinfo{volume}{B73}}, \bibinfo{pages}{115112}
  (\bibinfo{year}{2006}), \eprint{cond-mat/0511332}.

\bibitem[{\citenamefont{Lee}(2008{\natexlab{a}})}]{Lee:2008xs}
\bibinfo{author}{\bibfnamefont{D.}~\bibnamefont{Lee}}, \bibinfo{journal}{Phys.
  Rev.} \textbf{\bibinfo{volume}{C78}}, \bibinfo{pages}{024001}
  (\bibinfo{year}{2008}{\natexlab{a}}), \eprint{arXiv:0803.1280 [nucl-th]}.

\bibitem[{\citenamefont{Lee}(2008{\natexlab{b}})}]{Lee:2007a}
\bibinfo{author}{\bibfnamefont{D.}~\bibnamefont{Lee}}, \bibinfo{journal}{Eur.
  Phys. J.} \textbf{\bibinfo{volume}{A35}}, \bibinfo{pages}{171}
  (\bibinfo{year}{2008}{\natexlab{b}}), \eprint{arXiv:0704.3439
  [cond-mat.supr-con]}.

\bibitem[{\citenamefont{Lee et~al.}(2010)\citenamefont{Lee, Endres, Kaplan, and
  Nicholson}}]{Lee:2010qp}
\bibinfo{author}{\bibfnamefont{J.-W.} \bibnamefont{Lee}},
  \bibinfo{author}{\bibfnamefont{M.~G.} \bibnamefont{Endres}},
  \bibinfo{author}{\bibfnamefont{D.~B.} \bibnamefont{Kaplan}},
  \bibnamefont{and} \bibinfo{author}{\bibfnamefont{A.~N.}
  \bibnamefont{Nicholson}}, \bibinfo{journal}{PoS}
  \textbf{\bibinfo{volume}{LATTICE2010}}, \bibinfo{pages}{197}
  (\bibinfo{year}{2010}), \eprint{1011.3026}.

\bibitem[{\citenamefont{Endres et~al.}(2010)\citenamefont{Endres, Kaplan, Lee,
  and Nicholson}}]{Endres:2010sq}
\bibinfo{author}{\bibfnamefont{M.~G.} \bibnamefont{Endres}},
  \bibinfo{author}{\bibfnamefont{D.~B.} \bibnamefont{Kaplan}},
  \bibinfo{author}{\bibfnamefont{J.-W.} \bibnamefont{Lee}}, \bibnamefont{and}
  \bibinfo{author}{\bibfnamefont{A.~N.} \bibnamefont{Nicholson}},
  \bibinfo{journal}{PoS} \textbf{\bibinfo{volume}{LATTICE2010}},
  \bibinfo{pages}{182} (\bibinfo{year}{2010}), \eprint{1011.3089}.

\bibitem[{\citenamefont{L{\"u}scher}(1986)}]{Luscher:1986pf}
\bibinfo{author}{\bibfnamefont{M.}~\bibnamefont{L{\"u}scher}},
  \bibinfo{journal}{Commun. Math. Phys.} \textbf{\bibinfo{volume}{105}},
  \bibinfo{pages}{153} (\bibinfo{year}{1986}).

\bibitem[{\citenamefont{Beane et~al.}(2004)\citenamefont{Beane, Bedaque,
  Parreno, and Savage}}]{Beane:2003da}
\bibinfo{author}{\bibfnamefont{S.~R.} \bibnamefont{Beane}},
  \bibinfo{author}{\bibfnamefont{P.~F.} \bibnamefont{Bedaque}},
  \bibinfo{author}{\bibfnamefont{A.}~\bibnamefont{Parreno}}, \bibnamefont{and}
  \bibinfo{author}{\bibfnamefont{M.~J.} \bibnamefont{Savage}},
  \bibinfo{journal}{Phys. Lett.} \textbf{\bibinfo{volume}{B585}},
  \bibinfo{pages}{106} (\bibinfo{year}{2004}), \eprint{hep-lat/0312004}.

\bibitem[{\citenamefont{Seki and van Kolck}(2006)}]{Seki:2005ns}
\bibinfo{author}{\bibfnamefont{R.}~\bibnamefont{Seki}} \bibnamefont{and}
  \bibinfo{author}{\bibfnamefont{U.}~\bibnamefont{van Kolck}},
  \bibinfo{journal}{Phys. Rev.} \textbf{\bibinfo{volume}{C73}},
  \bibinfo{pages}{044006} (\bibinfo{year}{2006}), \eprint{nucl-th/0509094}.

\bibitem[{\citenamefont{Borasoy et~al.}(2007)\citenamefont{Borasoy, Epelbaum,
  Krebs, Lee, and Mei{\ss}ner}}]{Borasoy:2006qn}
\bibinfo{author}{\bibfnamefont{B.}~\bibnamefont{Borasoy}},
  \bibinfo{author}{\bibfnamefont{E.}~\bibnamefont{Epelbaum}},
  \bibinfo{author}{\bibfnamefont{H.}~\bibnamefont{Krebs}},
  \bibinfo{author}{\bibfnamefont{D.}~\bibnamefont{Lee}}, \bibnamefont{and}
  \bibinfo{author}{\bibfnamefont{U.-G.} \bibnamefont{Mei{\ss}ner}},
  \bibinfo{journal}{Eur. Phys. J.} \textbf{\bibinfo{volume}{A31}},
  \bibinfo{pages}{105} (\bibinfo{year}{2007}), \eprint{nucl-th/0611087}.

\bibitem[{\citenamefont{Lanczos}(1950)}]{Lanczos:1950}
\bibinfo{author}{\bibfnamefont{C.}~\bibnamefont{Lanczos}}, \bibinfo{journal}{J.
  Res. Nat. Bur. Stand.} \textbf{\bibinfo{volume}{45}}, \bibinfo{pages}{255}
  (\bibinfo{year}{1950}).

\bibitem[{\citenamefont{Lee and Thomson}(2007)}]{Lee:2007jd}
\bibinfo{author}{\bibfnamefont{D.}~\bibnamefont{Lee}} \bibnamefont{and}
  \bibinfo{author}{\bibfnamefont{R.}~\bibnamefont{Thomson}},
  \bibinfo{journal}{Phys. Rev.} \textbf{\bibinfo{volume}{C75}},
  \bibinfo{pages}{064003} (\bibinfo{year}{2007}), \eprint{nucl-th/0701048}.

\bibitem[{\citenamefont{Lee}(2007)}]{Lee:2006hr}
\bibinfo{author}{\bibfnamefont{D.}~\bibnamefont{Lee}}, \bibinfo{journal}{Phys.
  Rev.} \textbf{\bibinfo{volume}{B75}}, \bibinfo{pages}{134502}
  (\bibinfo{year}{2007}), \eprint{cond-mat/0606706}.

\bibitem[{\citenamefont{Lee}(2009)}]{Lee:2008fa}
\bibinfo{author}{\bibfnamefont{D.}~\bibnamefont{Lee}}, \bibinfo{journal}{Prog.
  Part. Nucl. Phys.} \textbf{\bibinfo{volume}{63}}, \bibinfo{pages}{117}
  (\bibinfo{year}{2009}), \eprint{arXiv:0804.3501 [nucl-th]}.

\bibitem[{\citenamefont{Creutz}(1988)}]{Creutz:1988wv}
\bibinfo{author}{\bibfnamefont{M.}~\bibnamefont{Creutz}},
  \bibinfo{journal}{Phys. Rev.} \textbf{\bibinfo{volume}{D38}},
  \bibinfo{pages}{1228} (\bibinfo{year}{1988}).

\bibitem[{\citenamefont{Creutz}(2000)}]{Creutz:1999zy}
\bibinfo{author}{\bibfnamefont{M.}~\bibnamefont{Creutz}},
  \bibinfo{journal}{Found. Phys.} \textbf{\bibinfo{volume}{30}},
  \bibinfo{pages}{487} (\bibinfo{year}{2000}), \eprint{hep-lat/9905024}.

\bibitem[{\citenamefont{Scalettar et~al.}(1986)\citenamefont{Scalettar,
  Scalapino, and Sugar}}]{Scalettar:1986uy}
\bibinfo{author}{\bibfnamefont{R.~T.} \bibnamefont{Scalettar}},
  \bibinfo{author}{\bibfnamefont{D.~J.} \bibnamefont{Scalapino}},
  \bibnamefont{and} \bibinfo{author}{\bibfnamefont{R.~L.} \bibnamefont{Sugar}},
  \bibinfo{journal}{Phys. Rev.} \textbf{\bibinfo{volume}{B34}},
  \bibinfo{pages}{7911} (\bibinfo{year}{1986}).

\bibitem[{\citenamefont{Gottlieb et~al.}(1987)\citenamefont{Gottlieb, Liu,
  Toussaint, Renken, and Sugar}}]{Gottlieb:1987mq}
\bibinfo{author}{\bibfnamefont{S.}~\bibnamefont{Gottlieb}},
  \bibinfo{author}{\bibfnamefont{W.}~\bibnamefont{Liu}},
  \bibinfo{author}{\bibfnamefont{D.}~\bibnamefont{Toussaint}},
  \bibinfo{author}{\bibfnamefont{R.~L.} \bibnamefont{Renken}},
  \bibnamefont{and} \bibinfo{author}{\bibfnamefont{R.~L.} \bibnamefont{Sugar}},
  \bibinfo{journal}{Phys. Rev.} \textbf{\bibinfo{volume}{D35}},
  \bibinfo{pages}{2531} (\bibinfo{year}{1987}).

\bibitem[{\citenamefont{Duane et~al.}(1987)\citenamefont{Duane, Kennedy,
  Pendleton, and Roweth}}]{Duane:1987de}
\bibinfo{author}{\bibfnamefont{S.}~\bibnamefont{Duane}},
  \bibinfo{author}{\bibfnamefont{A.~D.} \bibnamefont{Kennedy}},
  \bibinfo{author}{\bibfnamefont{B.~J.} \bibnamefont{Pendleton}},
  \bibnamefont{and} \bibinfo{author}{\bibfnamefont{D.}~\bibnamefont{Roweth}},
  \bibinfo{journal}{Phys. Lett.} \textbf{\bibinfo{volume}{B195}},
  \bibinfo{pages}{216} (\bibinfo{year}{1987}).

\bibitem[{\citenamefont{Foulkes et~al.}(2001)\citenamefont{Foulkes, Mitas,
  Needs, and Rajagopal}}]{Mitas:2001}
\bibinfo{author}{\bibfnamefont{W.}~\bibnamefont{Foulkes}},
  \bibinfo{author}{\bibfnamefont{L.}~\bibnamefont{Mitas}},
  \bibinfo{author}{\bibfnamefont{R.}~\bibnamefont{Needs}}, \bibnamefont{and}
  \bibinfo{author}{\bibfnamefont{G.}~\bibnamefont{Rajagopal}},
  \bibinfo{journal}{Rev. Mod. Phys.} \textbf{\bibinfo{volume}{73}},
  \bibinfo{pages}{33} (\bibinfo{year}{2001}).

\bibitem[{\citenamefont{Jastrow}(1955)}]{Jastrow:1955a}
\bibinfo{author}{\bibfnamefont{R.}~\bibnamefont{Jastrow}},
  \bibinfo{journal}{Phys. Rev.} \textbf{\bibinfo{volume}{98}},
  \bibinfo{pages}{1479} (\bibinfo{year}{1955}).

\bibitem[{\citenamefont{Umrigar et~al.}(1988)\citenamefont{Umrigar, Wilson, and
  Wilkins}}]{Umrigar:1988a}
\bibinfo{author}{\bibfnamefont{C.~J.} \bibnamefont{Umrigar}},
  \bibinfo{author}{\bibfnamefont{K.~G.} \bibnamefont{Wilson}},
  \bibnamefont{and} \bibinfo{author}{\bibfnamefont{J.~W.}
  \bibnamefont{Wilkins}}, \bibinfo{journal}{Phys. Rev. Lett.}
  \textbf{\bibinfo{volume}{60}}, \bibinfo{pages}{1719} (\bibinfo{year}{1988}).

\bibitem[{\citenamefont{Schmidt and Moskowitz}(1990)}]{Schmidt:1990a}
\bibinfo{author}{\bibfnamefont{K.~E.} \bibnamefont{Schmidt}} \bibnamefont{and}
  \bibinfo{author}{\bibfnamefont{J.~W.} \bibnamefont{Moskowitz}},
  \bibinfo{journal}{J. Chem. Phys.} \textbf{\bibinfo{volume}{93}},
  \bibinfo{pages}{4172} (\bibinfo{year}{1990}).

\bibitem[{\citenamefont{Mitas and Martin}(1994)}]{Mitas:1994a}
\bibinfo{author}{\bibfnamefont{L.}~\bibnamefont{Mitas}} \bibnamefont{and}
  \bibinfo{author}{\bibfnamefont{R.~M.} \bibnamefont{Martin}},
  \bibinfo{journal}{Phys. Rev. Lett.} \textbf{\bibinfo{volume}{72}},
  \bibinfo{pages}{2438} (\bibinfo{year}{1994}).

\bibitem[{\citenamefont{Grossman et~al.}(1995)\citenamefont{Grossman, Mitas,
  and Raghavachari}}]{Grossman:1995b}
\bibinfo{author}{\bibfnamefont{J.~C.} \bibnamefont{Grossman}},
  \bibinfo{author}{\bibfnamefont{L.}~\bibnamefont{Mitas}}, \bibnamefont{and}
  \bibinfo{author}{\bibfnamefont{K.}~\bibnamefont{Raghavachari}},
  \bibinfo{journal}{Phys. Rev. Lett.} \textbf{\bibinfo{volume}{75}},
  \bibinfo{pages}{3870} (\bibinfo{year}{1995}).

\bibitem[{\citenamefont{Bajdich and Mitas}(2009)}]{Bajdich:2009}
\bibinfo{author}{\bibfnamefont{M.}~\bibnamefont{Bajdich}} \bibnamefont{and}
  \bibinfo{author}{\bibfnamefont{L.}~\bibnamefont{Mitas}},
  \bibinfo{journal}{Acta Phys. Slovaca} \textbf{\bibinfo{volume}{59}},
  \bibinfo{pages}{81} (\bibinfo{year}{2009}).

\bibitem[{\citenamefont{Ceperley and Alder}(1984)}]{Ceperley:1984}
\bibinfo{author}{\bibfnamefont{D.}~\bibnamefont{Ceperley}} \bibnamefont{and}
  \bibinfo{author}{\bibfnamefont{B.}~\bibnamefont{Alder}}, \bibinfo{journal}{J.
  Chem. Phys.} \textbf{\bibinfo{volume}{81}}, \bibinfo{pages}{12}
  (\bibinfo{year}{1984}).

\bibitem[{\citenamefont{Li et~al.}(2011)\citenamefont{Li, Kolorenc, and
  Mitas}}]{Li:2011b}
\bibinfo{author}{\bibfnamefont{X.}~\bibnamefont{Li}},
  \bibinfo{author}{\bibfnamefont{J.}~\bibnamefont{Kolorenc}}, \bibnamefont{and}
  \bibinfo{author}{\bibfnamefont{L.}~\bibnamefont{Mitas}}
  (\bibinfo{year}{2011}), \eprint{arXiv:1105.1748v2 [cond-mat.quant-gas]}.

\end{thebibliography}

\end{document}